\documentclass{aa}  
\usepackage{natbib}
\bibpunct{(}{)}{;}{a}{}{,} 
\usepackage{graphicx}
\usepackage{txfonts}
\usepackage[pdfencoding=auto,psdextra]{hyperref}
\hypersetup{
    colorlinks=true,
    linkcolor=blue,
    filecolor=magenta,      
    urlcolor=blue,
    citecolor=blue
}
\usepackage{tabularx}
\usepackage{booktabs}

\begin{document}

   \title{Quiescent fractions in high-redshift galaxy groups reflect their hot-or-cold state of gas accretion }


   \author{Guillaume Elias
          \inst{1}\fnmsep\thanks{\email{guillaume.elias@cea.fr}}
          \and
          Emanuele Daddi\inst{1}
          \and
          Chiara D'Eugenio\inst{1}
          \and
          David Elbaz\inst{1}
          \and
          Maximilien Franco\inst{1}
          \and
          Fabrizio Gentile\inst{1}
          \and
          Raphael Gobat\inst{9}
          \and
          Sicen Guo\inst{1}
          \and
          Shuowen Jin\inst{7,8}
          \and
          Clotilde Laigle\inst{4}
          \and
          Shiying Lu\inst{1,5,6}
          \and
          Georgios E. Magdis\inst{7,8,10}
          \and
          Benjamin Magnelli\inst{1}
          \and
          Nikolaj B. Sillassen\inst{7,8}
          \and
          Veronica Strazzullo\inst{2,3}
          \and
          Maxime Tarrasse\inst{5,6}
          \and
          Tao Wang\inst{5,6}
          \and
          Luwenjia Zhou\inst{5,6}
          }

\institute{
Université Paris-Saclay, CNRS, CEA, Astrophysique, Instrumentation et Modélisation de Paris-Saclay, 91191, Gif-sur-Yvette, France.
\and
INAF-Osservatorio Astronomico di Trieste, Via Tiepolo 11, 34131, Trieste, Italy
\and
IFPU-Institute for Fundamental Physics of the Universe, Via Beirut 2, 34014, Trieste, Italy
\and
Institut d’Astrophysique de Paris, UMR 7095, CNRS, Sorbonne Université, 98 bis boulevard Arago, F-75014 Paris, France
\and
School of Astronomy and Space Science, Nanjing University, Nanjing 210093, China
\and
Key Laboratory of Modern Astronomy and Astrophysics (Nanjing University), Ministry of Education, Nanjing 210093, China
\and
Cosmic Dawn Center (DAWN), Denmark
\and
DTU Space, Technical University of Denmark, Elektrovej 327, DK-2800 Kgs. Lyngby, Denmark
\and
Instituto de Física, Pontificia Universidad Católica de Valparaíso, Casilla 4059, Valparaíso, Chile
\and
Niels Bohr Institute, University of Copenhagen, Jagtvej 128, DK-2200 Copenhagen, Denmark
}

   \date{Received XX; accepted YY}

 
  \abstract
  {Cold accretion and quenching are closely related aspects of galaxy evolution, as sustained gas supply is required to maintain star formation. High-redshift galaxy groups therefore provide a valuable laboratory for testing how the thermal state of accreting gas relates to the emergence of quiescence. We measure quiescent fractions in a sample of 16 spectroscopically confirmed galaxy groups at \(1.6<z<3.6\), spanning halo masses from \(10^{12.8}\,\mathrm{M_\odot}\) to \(10^{13.9}\,\mathrm{M_\odot}\), by fitting the SEDs of candidate member galaxies selected from the COSMOS2020 catalog and by using a membership-probability approach to estimate group quiescent fractions. We compare these quiescent fractions to the expected cold or hot accretion state of each halo and find evidence for a correlation: quiescent fractions reach \(\sim 50\%\) in groups in the hot-accretion regime and are consistent with 0\% in groups in the cold-accretion regime. In mature hot-accreting groups, massive quiescent galaxies are preferentially found in the inner regions (\(R<0.5R_{\rm vir}\)), with a $4.4\,\sigma$ excess relative to the outskirts. Most groups lack a clearly established BGG and instead show small stellar-mass gaps, typically $M_{*,1}/M_{*,2} \lesssim 3$, indicating that they remain in an active assembly phase rather than being dynamically evolved systems. Consistently, the stellar-mass excess of the dominant galaxy, measured relative to the SHMR expectation, does not predict the group quiescent fraction. Taken together, our results support a picture in which the cold-to-hot transition in gas accretion contributes to the onset of quiescence, possibly through inside-out starvation associated with filament disruption in shock-heated intra-group gas, and suggest that environment plays a greater role than internal processes in shaping the quiescent galaxy population in these structures.
  }


   \maketitle
%

\section{Introduction}



The death of galaxies is one of the most elusive problems in modern astrophysics \citep{man18, faber07}. It is a cornerstone of all facets of galaxy physics: star-formation, morphology, gas accretion, environment interactions, galaxy groups and large scale structure (LSS) formation. Many subjects of research contribute to answering the major question: why do galaxies stop forming stars? 

Internal processes have long been identified as a cause for the observed lower star formation efficiencies of galaxies, as well as for their lower abundances compared to what is allowed by cosmology: star-formation feedback is invoked at the low-mass end and active galactic nuclei (AGN) feedback at the high-mass end. These results mainly arose from simulation-based studies where these sub-resolution prescriptions describing these processes are essential to match observations \citep{croton06, somerville15}. However, recent comparisons between simulations and observations tend to show that AGN kinetic/jet-mode feedback models produce rapid, sometimes excessive quenching and under-predict star formation rates (SFRs) in certain mass/environment regimes \citep{donnari21, weinberger18, dave19, donnari19} and tests against circum-galactic medium (CGM) and X-ray constraints suggest AGN energetics in models require further refinement \citep{eckert24}.

AGN growth has been shown to be more pronounced at z>4 \citep{guetzoyan25} and black-hole mass to stellar mass ratios have been shown to be more important beyond redshift 6 \citep{merloni10, decarli18}. If AGN were more massive and active than their host galaxies in the past, how could their activity keep their host galaxies from growing, gaining mass and catching up to them later in their lifetime? Recent studies have suggested the idea that AGN feedback could be a phenomenon co-occurring with galaxy quenching, influenced by the same physics of accretion \citep{mullaney12, harrison17, kubo22}, meaning accretion physics would be responsible for both galaxy quenching and AGN feedback. Gas accretion physics are closely related to environmental effects since they are impacted by the location of the halo within the LSS and by its mass and evolutionary stage. Taken together, these findings suggest the need to investigate other potential channels to quench galaxies, such as those related to their environment.


Galaxies evolve in a complex environment formally defined by the cosmic web, forming along it and concentrating at its most dense points, the nodes or clusters, and fed gas flowing through filaments \citep{bond96, delapparent86}. The cosmic web is a key component of galaxy formation, galaxy properties such as their SFRs or morphologies depend on galaxies positions or alignments with large scale filaments and nodes \citep{laigle18, kraljic20, mondelin25}. Quenching is not simply related to internal processes but also to environmental effects such as ram pressure stripping, tidal interactions, galaxy harassment and mergers \citep{boselli06, vanderburg20, xu25}. These processes are especially relevant in crowded, overdense environments such as galaxy clusters which are predominantly populated by quiescent galaxies in the local Universe \citep{dressler80, balogh99}. At low redshift (z<1), mass-related and environment-related quenching processes are separable \citep{peng10} but this picture blurs at higher redshift, with mass- and environment-related processes showing cross-dependencies at higher redshifts \citep{darvish16, fossati17}. Studying overdense environments at high redshift is essential to understand quenching mechanisms since galaxies are subject to both internal and external quenching. The hierarchical nature of structure formation also implies that galaxies in these environments have started forming earlier and that they might be subject to group pre-processing \citep{zabludoff98}, meaning that their properties might be impacted by their site of formation without physics related to high-density environments.


High-redshift galaxy groups are ideal laboratories to study quenching at high redshift \citep{overzier16, chiang17}. Throughout the past decade, a growing number of individual discoveries and large observation programs have allowed to progressively build up a large sample of galaxy groups spanning a wide variety of redshifts and halo masses \citep{gobat11, strazzullo13, wang16, daddi21}, effectively moving the field from isolated case studies to the realm of systematic studies. Recent observations have shown that some $z \gtrsim 2$ overdense structures already host substantial populations of massive quiescent galaxies, indicating that galaxy quenching can be well underway before the emergence of fully virialized clusters \citep{mcconachie22, ito23}.


Galaxy groups are also defined by their position within the cosmic web, forming at its nodes and undergoing intense gas accretion. Direct imaging of filaments is scarce \citep{daddi21, martin15, bacon21, cantaloupo14} but indirect evidence of their role in shaping galaxy and cluster properties is becoming more prevalent. It has long been predicted that depending on the survival of cosmic web filaments crossing the virial radius of halos, the accretion of gas on clusters happens in either a hot or a cold mode \citep{dekel06}. These original prediction have been refined in more modern studies \citep{mandelker20} and further observations of correlations between galaxy and group halo give credibility to these theoretical predictions \citep{daddi22a, daddi22b}.

Rapid gas consumption timescales of a few hundred million years at cosmic noon \citep{coogan18, tacconi18} imply that gas renewal is essential to keep galaxies star-forming. As such, measuring quiescent fractions could be a good proxy for assessing the cold-or-hot state of the gas infalling on the halos and feeding galaxies. The state of the gas and the way it deposits into the halos also leave an imprint on the distribution of quiescent/star-forming galaxies in the mass/distance plane of halos. It could allow to better differentiate scenarios in which accretion feeds mostly the satellite galaxies or the central ones depending on the distance and mass of quiescent galaxies in the halos and overall constrain quenching mechanisms \citep{vanderburg20, fossati17, wetzel13}.

In the absence of cooling flows at the center of galaxy groups, cold gas renewal happens only through filamentary accretion through the cosmic web, hence the state of the gas being accreted should have a crucial role on group properties and especially on their quiescent fractions. Works such as \cite{overzier16} first introduced the idea that these two concepts should be related (Figure~19 of their work) and \cite{behroozi19} investigated the relation between the two from the perspective of simulations (Figure~15 of their work). The exact shape of the transition from cold-in-hot to hot accretion in the halo-mass -- redshift parameter space is subject to debate but its principle is widely accepted and the idea that it should reflect on galaxy star-formation is gaining momentum. {For galaxy groups accreting gas in cold mode, this gas should remain readily available to sustain star formation, and lower quiescent fractions are therefore expected than in systems where cold inflow is suppressed. Testing this expectation observationally requires relating the quiescent populations of galaxy groups to their expected accretion state in the halo-mass--redshift plane.}

{An independent large-sample confirmation of this global trend was reported by \citet{sutanto26}, using 207 photometrically selected overdensities identified with JWST. They find higher quiescent fractions and lower average sSFRs in systems occupying the hot-accretion regime than in the cold-in-hot regime, as well as a dependence of these quantities on distance from the predicted cold-stream threshold. Their substantially larger sample therefore confirms statistically the relation between galaxy star-formation activity and the expected accretion regime. In this work, we investigate the physical origin of this relation using 16 spectroscopically confirmed galaxy groups, examining which galaxies are quenched, where they reside within the groups, whether alternative explanations related to internal galaxy properties can account for the observed behavior, and how the results depend on the main sources of observational and methodological uncertainty.}

{Our groups span redshifts from $1.6$ to $3.6$ and halo masses from $10^{12.8}\,M_\odot$ to $10^{13.9}\,M_\odot$. We analyze galaxy membership and stellar populations using the COSMOS2020 catalog \citep{cosmos20} and FAST++, deriving physical properties to classify galaxies as quiescent or star-forming. We then relate the resulting quiescent fractions to the properties of their host halos and investigate the distributions of quiescent and star-forming galaxies in stellar mass and group-centric distance.}

Section \ref{section:methods} details the methodology, namely group selection, galaxy membership determination, SED fitting routine and quiescent classification. Section \ref{section:results} details our results, showcasing the red sequence of galaxies in our groups, analyzing quiescent fraction correlations with host properties and mass/distance distributions. Finally, Section \ref{section:discussion} puts our results in perspective with the current state of the art in galaxy quenching and high-z galaxy group science. 

In the following, we assume a $\Lambda$CDM cosmology with 
$H_{0} = 67.4\,\mathrm{km\,s^{-1}\,Mpc^{-1}}$, 
$\Omega_{\mathrm{M}} = 0.315$, 
and $\Omega_{\Lambda} = 0.685$ \citep{planck}.

\section{Methodology} \label{section:methods}

\subsection{Group selection}

\begin{table*}
\caption{Table of physical properties of the groups analyzed in this study. For groups from HPC1001 to SBC6, halos masses and virial radii are taken from \cite{sillassen24}. For groups from CC-0958 to RO-1001, they are taken from \cite{daddi22b}. For the two satellite groups ('-Sat'), they are taken from \cite{guo25}. {The derivation of these quantities and their uncertainties is summarized in the text.}}
\label{tab:group_prop}
\centering
\begin{tabular}{c c c c c c}  
\hline\hline
ID & z & R.A. (°) & Dec (°) & $\rm{log}_{10}(M_h/M_\odot)$ & $\rm{R_{vir} (pkpc) }$ \\ 
\hline
HPC1001 & 3.610 & 150.4656 & 2.6360 & 13.3 & 181 \\ 
\hline
SBCX1 & 2.422 & 150.3480 & 2.7611 & 13.6 & 307 \\ 
\hline
SBCX3 & 3.031 & 150.3112 & 2.4511 & 12.8 & 141 \\ 
\hline
SBCX4 & 2.646 & 150.7508 & 2.4129 & 13.3 & 229 \\ 
\hline
SBCX7 & 2.415 & 149.9890 & 1.7976 & 13.6 & 307 \\ 
\hline
SBC3 & 2.365 & 150.7225 & 2.6963 & 13.5 & 289 \\ 
\hline
SBC4 & 1.644 & 150.0367 & 2.2178 & 13.6 & 397 \\ 
\hline
SBC6 & 2.323 & 149.7053 & 2.2159 & 13.4 & 271 \\ 
\hline
CC-0958 & 2.515 & 149.7206 & 1.9674 & 13.6 & 299 \\ 
\hline
CLJ1001 & 2.501 & 150.2382 & 2.3356 & 13.9 & 378 \\ 
\hline
FVX-LAB & 2.194 & 149.6763 & 2.0109 & 13.0 & 207 \\ 
\hline 
RO-0958 & 3.295 &  149.5824   & 2.6028 & 12.9 & 143 \\ 
\hline
RO-0959 & 3.096 & 149.9978 & 2.5782 & 12.8 & 138 \\ 
\hline
RO-0959-Sat & 3.092 & 149.9876 & 2.5769 & 12.8 & 139 \\ 
\hline
RO-1001 & 2.915 & 150.3460 & 2.3346 & 13.6& 268 \\ 
\hline
RO-1001-Sat & 2.920 & 150.3510 & 2.3438 & 13.2 & 197 \\

\end{tabular}

\end{table*}

The galaxy groups presented in this study were detected as part of the NICE observation program \citep{zhou24, sillassen24, zhou25}, a large NOEMA and ALMA observation program targeting high redshift over-densities. Galaxy groups were originally selected as overdensities of massive IRAC--selected high–redshift galaxies at the $5\sigma$ level, coincident with Herschel/SPIRE $350\,\mu\mathrm{m}$--peaking, IR--bright sources indicative of intense, collective star formation within massive halos in the redshift range $2<z<4$. See \cite{zhou24} for an exhaustive description of the selection criteria of the targets.

Out of the 25 candidates selected in this fashion, 8 of them were spectroscopically confirmed in the COSMOS field. They were characterized in \cite{sillassen24} and table \ref{tab:group_prop} gives a summary of their findings that we use in this study. The gas content of these groups as well as of those detected outside of the COSMOS field was studied in \cite{zhou25}.

To complement this homogeneously selected sample, we also include in this study 6 additional groups studied as part of previous work focused on observational evidence for the cold-to-hot accretion regime transition from \cite{daddi22a}, as well as two satellite groups identified in \cite{guo25}, all of which are located within the COSMOS field. The inclusion of these groups allows us to probe more extensively the parameter space of high-redshift galaxy groups and brings the total number of groups in this study to 16. Altogether, these groups' redshifts span 1.6 to 3.6 and their halo masses range from $10^{12.8} \rm{M_\odot}$ to $10^{13.9} \rm{M_\odot}$. {We summarize below how the halo masses and virial radii entering our analysis were determined.}

{The halo masses used throughout this work come from the dedicated characterization of these structures in the studies from which the group sample was assembled. For the eight NICE groups, \citet{sillassen24} estimated halo masses using six estimators spanning three main approaches. Their best estimate combines three of these: (i) the total stellar mass of candidate members, corrected for field contamination and incompleteness and converted to halo mass using the redshift-dependent central+satellite SHMR of \citet{shmr}; (ii) the projected galaxy overdensity relative to the field, converted to a matter overdensity using the mass-dependent clustering bias; and (iii) a fit of the projected stellar-mass density profile with a NFW halo model. These three estimates agree with a scatter of $\simeq0.2$--$0.3$ dex and systematic offsets below $0.2$ dex. The adopted halo mass is their average, with a conservative uncertainty of $0.3$ dex \citep{sillassen24}. The halo masses of the two satellite groups from \citet{guo25} were obtained following the same methodology.}

{For the six groups drawn from \citet{daddi22b}, the halo masses were primarily estimated from their integrated stellar content. Spectroscopically confirmed members and galaxies with consistent photometric redshifts were used to measure the integrated stellar mass within the extent of each structure; this mass was corrected below the stellar-mass completeness limit using the stellar mass functions of \citet{muzzin13} and converted to $M_{200}$ using the calibration of \citet{vanderburg14}. Where available, these stellar-mass-based estimates were checked against independent X-ray, SZ, or previously available dynamical constraints. In particular, \citet{daddi22b} estimate uncertainties of $\lesssim0.2$ dex for the more massive systems for which such independent constraints are available, while the lower-mass systems have uncertainties of approximately $0.3$--$0.4$ dex. Throughout this work we define $R_{\rm vir}\equiv R_{200}$. For the NICE systems we adopt the published radii derived from their best halo masses using \citet{goerdt10} and for the remaining systems we calculate $R_{200}$ self-consistently from the adopted $M_{200}$ and redshift using the same formula. The uncertainty on $R_{\rm vir}$ is not independent of that on $M_h$: because $R_{200}\propto M_{200}^{1/3}$ at fixed redshift, a $0.3$ dex uncertainty in halo mass corresponds to an uncertainty of approximately $25\%$ in virial radius. We discuss the sensitivity of our results to the adopted radial selection in Section~\ref{section:limitations}.}

\subsection{Galaxy sample selection/contamination}\label{section:contamination}

\begin{figure}
   \centering
    \includegraphics[scale=0.63]{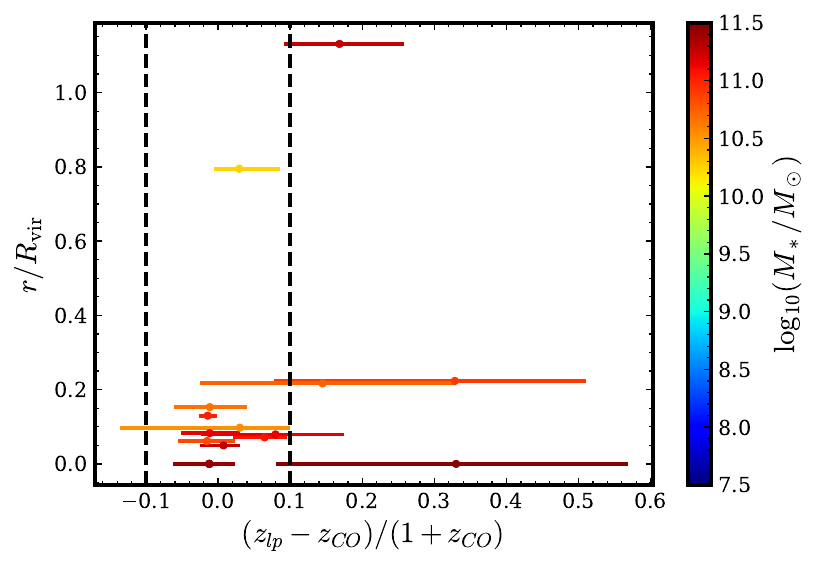}
    \includegraphics[scale=0.63]{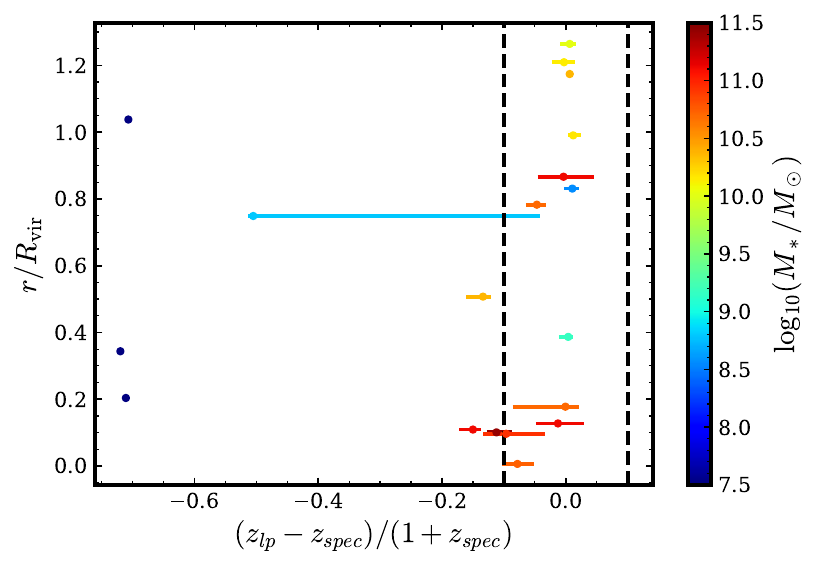}
   \caption{Comparison between photo z and spec z. The x-axis is $(photo_z - spec_z)/(1+spec_z)$ and y-axis is distance from the galaxy to its host group expressed as a fraction of the group's virial radius. Colors correspond to the masses of galaxies. \textit{Top panel} shows comparison to spectroscopic redshifts obtained using the NICE observations \cite{khostovan26}. \textit{Bottom panel} shows comparison to spectroscopic redshifts from the COSMOS Spectroscopic Redshift Compilation \citep{khostovan26}. {For a few galaxies, the photometric-redshift uncertainties are smaller than the marker size and the corresponding horizontal error bars are therefore not visible.}}
    \label{fig:photo_z}
    \end{figure}

Only a handful of galaxies are spectroscopically confirmed in each group, hence ancillary data is necessary to characterize the population of group members. To that end, we use the COSMOS2020 catalog \citep{cosmos20} containing deep photometry and physical properties of the galaxies in the COSMOS field. We acknowledge the recent release of the COSMOS 2025 catalog containing the JWST observations of the field, but since half of the groups in our sample are not covered by these observations, we decided to keep this dataset for a forthcoming study that will focus on checking the validity of our analysis in light of the new insights provided by the JWST.

{Throughout this work, $z_{\rm phot}$ denotes the median of the LePhare photometric-redshift probability distribution provided in COSMOS2020, corresponding to the \texttt{lp\_zPDF} quantity \citep{cosmos20}. We use this value as a point estimate for the initial redshift pre-selection.}

We select galaxies based on the groups redshifts and virial radii. A first pre-selection is performed by taking galaxies within $1 R_{vir}$ of the group center and whose photometric redshift falls within 10\% of the group redshift, or in other words that verify: 

\begin{equation}
\left|z_{phot} - z_{group}\right|< 0.1 \cdot \left(1+z_{group} \right)
\end{equation}

{This is deliberately an initial, relatively broad selection rather than a secure membership criterion; the subsequent selection and weighting procedure explicitly accounts statistically for contamination by field galaxies.}

\cite{cosmos20} find that for the faintest sources, photometric redshifts have an accuracy of about 5\% and the fraction of outliers (sources for which $\left|z_{phot} - z_{spec}\right|> 0.15 \cdot \left(1+z_{spec} \right)$) can reach up to 25\%. Given these numbers, selecting sources with photometric redshift falling within 10\% of the group redshift seems like a fair approach. Checking the photometric redshifts of galaxies for which we have spectroscopic information shows that this is a fair approach, most of their photometric redshifts falling within 10\% of their spectroscopic one (shown Figure~\ref{fig:photo_z}). The ones for which this is not the case all have large photometric redshift uncertainties, motivating an additional cut to remove galaxies with similar errors that would thus likely pollute our sample because of inaccurate redshift estimation. 

We also apply the mass completeness cut for quiescent galaxies described in section 6.2 of \cite{cosmos20} to ensure that our sample is mass complete. This cut is redshift dependent and varies almost linearly from $\rm{log_{10} (M_*/M_\odot)} = 9$ at redshift 1.6 to $\rm{log_{10} (M_*/M_\odot)} = 9.6$ at redshift 3.6. \\

Once this preselection is done, we refine this sample to have a handle on the contamination from field galaxies. This is done in two alternative ways to achieve the same purpose. For both, we start by estimating the average number of field galaxies per unit area at a given mass using the same redshift selection as described in the previous paragraph to estimate the density of contaminants in the field for each group. Then, for the first method, we perform a simple 2-dimensional stellar mass and distance to the group cut giving a sample with 10\% of contaminants, meaning where the lower $M_{*, \rm inf}$ bound and upper $r_{\rm max}$ bound give a number of contaminants $N_{\rm contaminants}$ such that its ratio with the number of galaxies $N_{\rm gal}$ with the same mass/radius cuts among the preselected ones is $\frac{N_{\rm contaminants}}{N_{\rm gal}} = 10\%$. As several $M_{*_{\rm inf}}$ and $r_{\rm max}$ combinations verify the same criteria, we choose the one that maximizes the number of galaxies $N_{gal}$ in the sample. This approach is the most straightforward way to select a sub-sample of the parameter space by having a handle on the field contamination, but it has the downside of excluding massive galaxies falling far away from the groups and low-mass galaxies that are close to it.

To palliate this issue, we introduce a second way of selecting galaxies, based on selecting areas of the parameter space where field contamination is lower preferentially. To do so, we scan the parameter space continuously, starting from highest $M_*$ and closest distance to the groups where contamination is the lowest, and scanning the space following the gradient of number of field contaminants. The contamination fraction is recomputed at each step, typically it decreases if the next cell of the parameter space contains a galaxy (since $N_{\rm gal}$ increases by $1$ and $N_{\rm contaminants}$ only by a fractional amount) and increases otherwise. The scanning stops when a fraction of 10\% is reached. This method also has the advantage of having less degrees of freedom than the previous one that gives several combinations of mass/distance cuts verifying the 10\% condition. Another advantage of this second method is that it allows to associate individual contamination fractions to galaxies, rather than only the bulk value of the sample for the first method.

\begin{figure}
   \centering
   \includegraphics[scale=0.57]{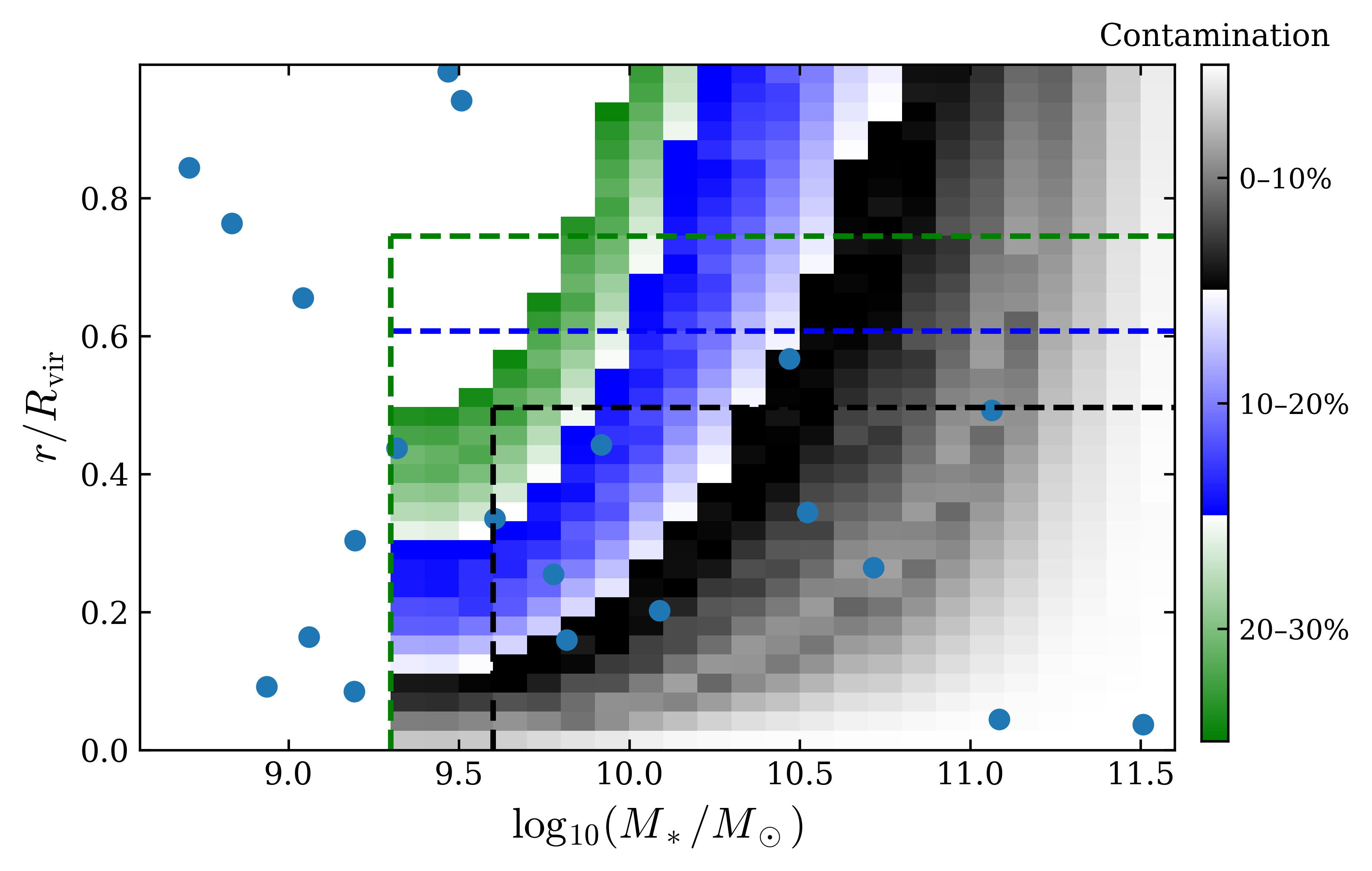}
   \caption{Figure showing the level of contamination throughout the parameter space for one of the groups in our study (CC-0958). Blue points correspond to candidate member galaxies. Background colors indicate the level of contamination according to the method where it is computed by scanning the parameter space from most to least contaminated. Dashed boxes correspond to the selections in parameter space according to the single mass/radius cut method (black corresponds to 10\%, blue to 20\% and gray to 30\%). The global mass completeness cut applied for this group is at $\rm{log_{10} (M_*/M_\odot)} = 9.3$.}
    \label{fig:contaminant}
    \end{figure}

Figure \ref{fig:contaminant} shows the areas selected in the parameter space according to both methods. It is visible that the second one allows to include massive galaxies sitting further away and lower mass galaxies close to the groups than what the first method allows. This is essential as these structures are in the process of assembling and as such, it is to be expected that massive galaxies do not necessarily sit at the center of the groups and some of the physics at play would be missed otherwise.

This procedures were described for the case of a sample containing 10\% of contaminants but can very well be applied with higher thresholds of 20\% or 30\%. This is important to test the robustness of observed trends to the presence of contaminants since the threshold of contaminants is somewhat arbitrary. The final samples we use at a given contamination percentage consists of the union of the samples given by the two methods with the same percentage, given that each method has its advantages and drawbacks.

\subsection{SED fitting} \label{section:sed}

Once a sample of galaxies has been selected, we fit their spectral energy distribution (SED) in order to extract and analyze their physical properties. To that end, we use the code FAST++\footnote{\url{https://github.com/cschreib/fastpp}}, an improved and faster version of the code FAST \citep{kriek09}. FAST++ operates $\chi^2$ minimization to find the best-fitting solution among the set of models generated by the parameter grid. We fit the observations to models generated with the Maraston 2013 \citep[M13,][]{m13} stellar templates which include accurate descriptions of thermally-pulsating asymptotic giant branch (tp-AGB) stars that drive a large part of the emission of evolved galaxies \citep{lu24}. We also fit our data to the more widely used Bruzual and Charlot 2003 templates \citep[BC03,][]{bc03} in order to allow for easier comparison between our results and the literature. {Nebular emission lines are not included in either set of stellar population models used for the fitting. \cite{lebail24} found in a similar sample of quiescent galaxies at cosmic noon that including them lead to higher $\chi^2$ without significantly affecting the derived physical properties.}

We use exponentially declining star-formation histories (SFH) since we are interested in the quiescent population of galaxies in the systems we study. They are of the form:
\begin{equation}
    \textrm{SFH} \propto \textrm{e}^{-\frac{t}{\tau}},
\end{equation}
 
where $t$ is the age of the stellar population of the galaxy and $\tau$ is the characteristic timescale of the exponential decline. These SFH offer several advantages: keeping the number of free parameters low and allowing a simple interpretation of the value of $t/\tau$, which can also be used as a summary statistic to classify galaxies as quiescent or star forming \citep{lebail24}. They also have the benefit of working both for star-forming galaxies, where $\frac{t}{\tau}\ll 1 $ and hence simplifies as $\textrm{SFH} \propto cst$ constant star-formation history, and for quiescent galaxies that either are in the process of stopping star formation ($\frac{t}{\tau} \gtrsim 1$) or have completely shut down star-formation ($\frac{t}{\tau} \gg 1$). The grid of parameters for $t$ and $\tau$ is given in table~\ref{tab:SED}.

We use the \cite{calzetti00} dust law. Newer dust laws have been advanced to model high-redshift galaxies, accounting for dust distributions affected by supernovae for example \citep{mckinney25}, but since dust obscuration is not a central focus of this work and to allow for easier comparison to the literature, we chose a dust law more commonly used. The values for $A_V$ explored in the fits are given in table~\ref{tab:SED}.

We do not fit the redshift of the galaxies. Instead, once a galaxy is associated to a group, we use the group redshift as the galaxy redshift.


To sample the galaxies' SEDs from the UV to the mid-infrared, we use the deepest available photometry in the COSMOS2020 catalog, namely CFHT~$u$, Subaru/Suprime-Cam~($B$, $V$), HSC~($g$, $r$, $i$, $z$, $y$), UltraVISTA~($Y$, $J$, $H$, $K_{\mathrm{s}}$), and \textit{Spitzer}/IRAC channels~1 and~2. We use the 2" aperture photometry for all bands. The photometry of all bands is corrected for Milky-Way attenuation and the photometry issued from of all ground-based facilities is corrected for atmospheric broadening following the procedure described in the files accompanying the catalog. We follow the same approach as described in Section~5.1 of \cite{cosmos20} to increase the errors on the fluxes used to account for errors in color modeling (adding in quadrature increasingly high error values to the magnitude errors, from 0.02~mag in the optical to 0.1~mag for the longest wavelengths).

\begin{table}[t]
\centering
\caption{Table of the grid of fitted parameters.}
\label{tab:SED}
\begin{tabular}{c c c c} 
\hline\hline
Parameter & min & max & step \\ 
\hline
$\rm A_v$ & 0 & 5 & 0.02 \\ 
\hline
${\rm log}_{10} (\tau)$ & 8 & 11 & 0.1 \\
\hline
${\rm log}_{10} (t)$ & 8 & 10 & 0.1 \\

\end{tabular}
\end{table}

For each best-fitting model, the code outputs various physical quantities such as stellar mass, star-formation rate (instantaneous), rest-frame colors and of course the best-fitting values for the fitted parameters. For each parameter, errors within a given uncertainty range can be derived directly from the $\chi^2$ grid of parameters or from Monte-Carlo (MC) simulations. We choose to use the MC errors as they allow to avoid assuming a $\chi^2$ distribution to derive them from the grid of results. 

To make sure the results of our fits do not depend on the coarseness of the grid of parameters, we run additional fits with a grid twice as thin as the original one and check that the best fitting values do not deviate from the original run. The new table of parameters fitted as well as figures illustrating the offsets can be found in Appendix \ref{appendix}, as well as figures comparing the best fitting values across the Bruzual \& Charlot and Maraston templates runs.

\subsection{Classification}

\begin{figure}
   \centering
   \includegraphics[scale=0.63]{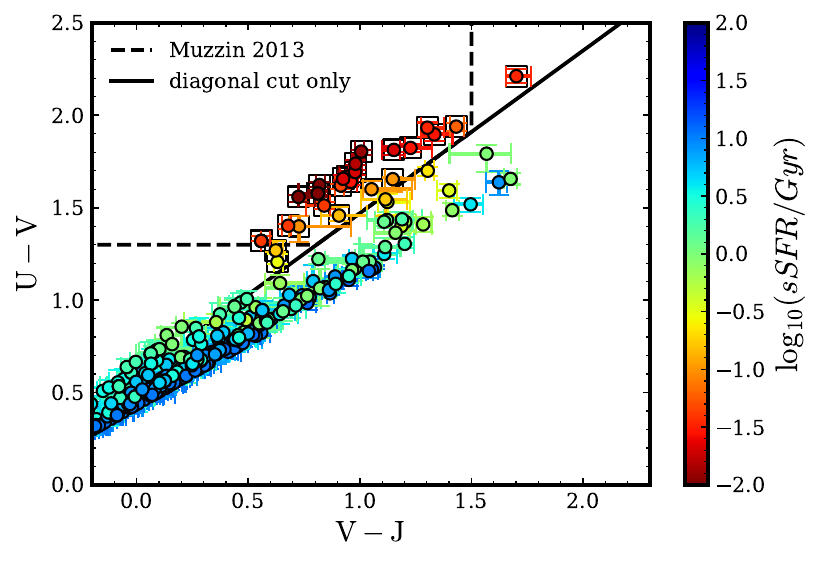}
   \caption{UVJ diagram of the galaxies in our sample. Square markers indicate quiescent galaxies.}
    \label{fig:red}
    \end{figure}

We classify galaxies as quiescent or star-forming according to several indicators. Similar classification approaches have been adopted in previous studies \cite[see for example][]{lebail24} that we follow broadly, albeit with a few minor changes. A first indicator is the star-forming main sequence (SFMS or MS thereafter), commonly referred to as the main sequence \citep{daddi07, elbaz07, noeske07}. The SFMS is the tight correlation ($\approx 0.3 \textrm{dex}$ scatter) between a galaxy's SFR and its stellar mass. Its existence is generally interpreted as evidence that a large fraction of star-forming galaxies evolve through relatively steady growth, although individual systems may depart from that behavior \citep{arangotoro2025}. The SFMS has been extensively studied, including its evolution with redshift, allowing its use as a ruler to measure a galaxy's SFR and compare it to the "normal" rate corresponding to its mass and redshift. Although using the SFMS as a ruler to define quiescent galaxies is somewhat recursive since one has to define star-forming galaxies to measure the SFMS in the first place, it still makes sense since its shape and properties are mostly unaffected by the choice of definition \citep{pearson23}. We use the \cite{schreiber15} redshift dependent parametrization of the MS for our study. If a galaxy's SFR falls below the SFMS by more then 2 times its scatter, we classify it as quiescent. This is the same threshold as assumed by \cite{lebail24}:

\begin{equation}
    \Delta MS \equiv \textrm{log}_{10}\left(\frac{\textrm{SFR}} {\textrm{SFR}_{\textrm{MS}} (\textrm{M}_*, z)} \right) < -0.6 
\end{equation}

A second criterion is how evolved the stellar population of a galaxy is. This is quantified by the value of the ratio of $\frac{t}{\tau}$, as detailed in section \ref{section:sed} and we chose to classify galaxies as quenched if they verify $\frac{t}{\tau}>3$ which is a rather conservative criteria given that $\frac{t}{\tau}=1$ is the turnover value between the star-forming stage and the quiescent one and that when $\frac{t}{\tau}=3$, the SFR reaches a value that is only 5\% of what is was at that turnover point which is residual. \cite{lebail24} chose a cut at $\frac{t}{\tau}=1$ but for our dataset, choosing this value proved to not be constraining, most galaxies verifying the other two criteria having $\frac{t}{\tau}$ values well over $1$.

\begin{figure}
   \centering
   \includegraphics[scale=0.63]{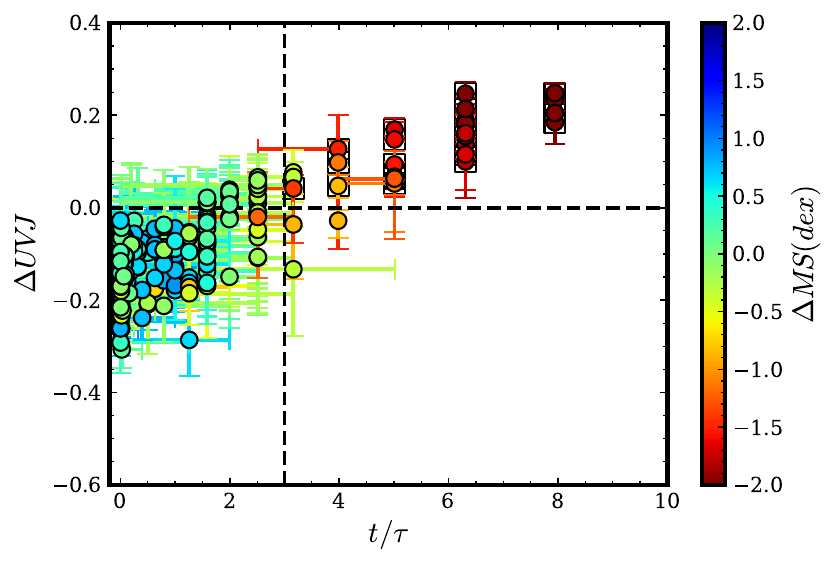}
   \caption{Distribution of all the values of the criteria used to classify galaxies as quiescent. Square markers indicate quiescent galaxies, i.e. galaxies for which best fit values match the criteria described in the text. $\Delta UVJ$ is the perpendicular distance to the border in the UVJ diagram, taken positive when in the quiescent region of the diagram.}
    \label{fig:criteria}
    \end{figure}

Finally, we use the empirical criteria based on rest-frame UVJ colors.
The UVJ diagram is a powerful tool to identify quiescent galaxies since quiescent gaalaxies and star-forming galaxies sit in different spaces of the diagram. Recent works have allowed to identify the paths and locations of galaxies in the diagram depending on the quenching timescale and on the properties of galaxies \citep{belli19, zick18}. 
Following works like \cite{belli19} and \cite{baker25}, we use only the diagonal cut of \cite{muzzin13} to define quiescent galaxies, given that the areas under the horizontal cut on the left corresponds to fast-quenched galaxies and the area on the right of the vertical cut corresponds to dusty-quiescent galaxies. Our sample contains few galaxies in the area corresponding to the extended region of the diagonal cut so choosing to use only the diagonal cut has little impact on our selection of quiescent galaxies, as seen in Figure~\ref{fig:red}.

Figure \ref{fig:criteria} shows the distribution and correlations of all the criteria described above. They span continuously the sampled parameter space, underscoring that quiescence is an undergoing process in these systems. The selection criteria described above are in agreement in the vast majority of cases, showing that our selection method is robust and self-consistent.

\subsection{Quiescent fraction estimation and uncertainty propagation}\label{section:q_frac}

For each galaxy in a given group we estimated the probability of being quiescent from the posterior samples produced by the SED fitting procedure. At each MC realization we evaluated the set of physical criteria defining quiescence: $\frac{t}{\tau}$ ratio, distance to the SFMS ($\Delta MS$) and distance to the border in the UVJ diagram ($\Delta UVJ$). This yields, for each galaxy $i$, a binary classification at each posterior sample, from which we define the quiescent probability
\begin{equation}
p_i = P_i(\mathrm{Q} \mid \mathrm{data}),
\end{equation}
equal to the fraction of posterior samples satisfying all quiescence criteria simultaneously.

The quiescent fraction of each group was then estimated by propagating these probabilities through Monte Carlo realizations. In each realization, the state of each galaxy was drawn from a Bernoulli distribution,
\begin{equation}
Z_i \sim \mathrm{Bernoulli}(p_i),
\end{equation}
where $Z_i = 1$ for a quiescent galaxy and $Z_i = 0$ otherwise. The group quiescent fraction was computed as a weighted mean
\begin{equation}
f_q = \frac{\sum_i w_i Z_i}{\sum_i w_i},
\end{equation}
where $w_i$ is the weight assigned to galaxy $i$. These weights account for the estimated statistical contamination from field galaxies as a function of stellar mass and projected distance from the group center as obtained through the second method described in Section~\ref{section:contamination}. This procedure naturally accounts for both measurement uncertainties in the galaxy classifications and statistical field contamination corrections.

Repeating this procedure over many realizations produces a posterior distribution for the group quiescent fraction. When this distribution is well defined, we adopt the median as the central estimate and the 16th and 84th percentiles as the corresponding confidence interval.

In some groups, however, all galaxies are effectively classified deterministically in the SED posterior (i.e.\ $p_i \simeq 0$ or $1$). In these cases the Monte Carlo distribution of $f_q$ becomes artificially discrete and may collapse to a single value, failing to reflect the finite-number uncertainty associated with the small number of galaxies in the group. To account for this effect, we adopt a separate treatment for these limiting cases.

Let $N$ be the number of galaxies contributing to the group measurement and $f_q$ the weighted quiescent fraction. We define the expected number of quiescent and star-forming galaxies as
\begin{equation}
N_q = f_q N, \qquad N_{\mathrm{SF}} = (1-f_q)N.
\end{equation}
If $N_q < 0.5$, the group is treated as effectively containing no quiescent galaxies and we report a one-sided upper limit
\begin{equation}
f_{q,\mathrm{up}} = 1 - \alpha^{1/N},
\end{equation}
where $\alpha=0.16$ corresponds approximately to a one-sided $1\sigma$ confidence level. Similarly, if $N_{\mathrm{SF}} < 0.5$, the group is treated as containing only quiescent galaxies and a corresponding lower limit is reported. 

In the intermediate case where both $N_q$ and $N_{\mathrm{SF}}$ exceed this threshold, the uncertainty is estimated from the standard binomial approximation \citep{cameron11},
\begin{equation}
\sigma_{f_q} = \sqrt{\frac{f_q (1-f_q)}{N}}.
\end{equation}

This hybrid approach preserves the full propagation of classification uncertainties when the quiescent probabilities are non-trivial, while ensuring a realistic treatment of small-number statistics in groups where galaxy classifications become effectively deterministic.

\section{Results} \label{section:results}

Having developed this pipeline to select galaxy samples for each group, deriving their physical properties and classifying their star-formation activity, and we can now analyze how the properties and distributions of quiescent galaxies relate to cold accretion theory and the large scale structure formation.

\subsection{Quiescent fractions and cold accretion theory}\label{results:cold_accretion}

\begin{figure*}[t]
   \centering
   \includegraphics[scale=0.63]{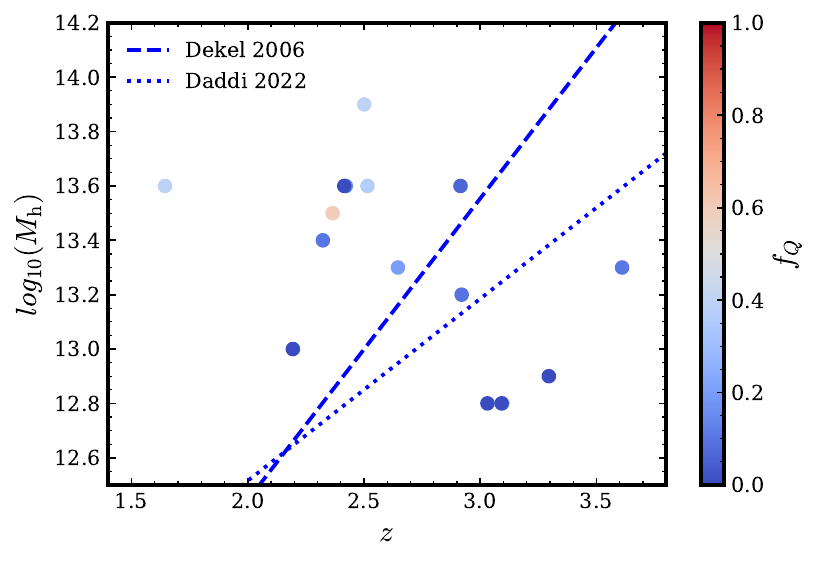} 
   \includegraphics[scale=0.63]{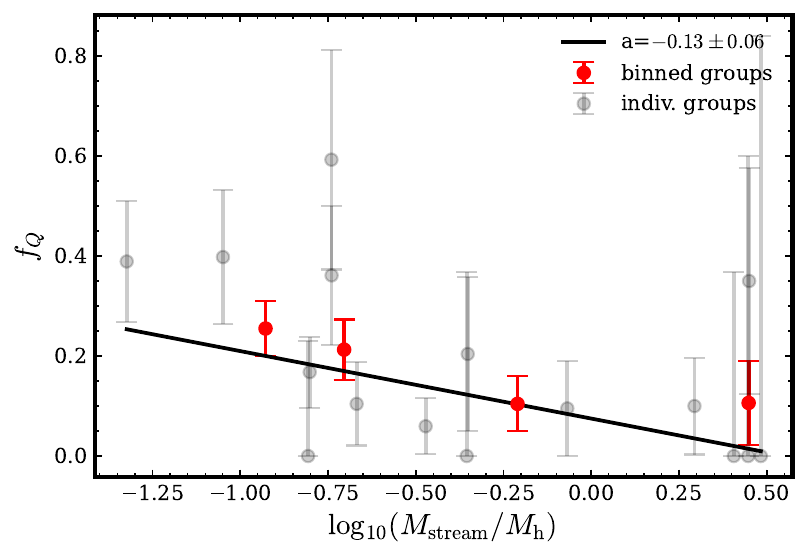}
   \caption{\textit{Left}: Groups distribution in the halo mass--redshift diagram, color-coded by the value of their quiescent fractions. The dashed and dotted lines represent the delimitation between the hot and cold-in-hot accretion regimes, as defined by \cite{dekel06} and \cite{daddi22b} respectively. \textit{Right}: Quiescent fractions as a function of $M_{\rm stream}/M_{\rm h}$, using the \cite{daddi22b} definition of $M_{\rm stream}$. Gray points correspond to the values for individual groups while the red points are obtained by stacking groups in bins of equals number of groups. Solid line represents the best-fit of individual groups using an affine function.}
    \label{fig:qfrac_mstream}
    \end{figure*}

Investigating quiescent fractions and their dependence with the state of gas accretion on galaxy groups is the pinnacle of this work. Given that central cooling in local massive halos is strongly suppressed relative to classical cooling-flow expectations \citep[the so-called "cooling flow problem",][]{mcdonald18}, and that cluster core gas densities appear not to increase significantly with redshift \citep{mcdonald17}, cooling from the group/cluster center alone is unlikely to provide a substantially larger fuel supply for star formation at high redshift. This makes ongoing gas accretion onto galaxies an important ingredient in sustaining star formation in forming groups.

{The shutdown of cold gas inflow should be reflected in the quiescent population of galaxy groups, making it important to compare their quiescent fractions with the expected state of gas accretion. A connection between these quantities was suggested by \citet{overzier16} and explored from the perspective of simulations by \citet{behroozi19}. Here we test this relation observationally using spectroscopically confirmed galaxy groups. }

{We first examine how the quiescent fraction $f_{\rm Q}$ of each group relates to its position in the halo mass--redshift plane, as shown in Figure~\ref{fig:qfrac_mstream}. The measured quiescent fractions follow the orientation expected from the theoretical accretion regimes: they are low in the cold-in-hot region and become larger farther into the hot-accretion regime. \citet{sutanto26} independently recover the same global behavior in a larger sample of photometrically selected JWST overdensities, providing a large-statistics confirmation of the trend seen in our spectroscopically confirmed systems.}

To better illustrate this trend, we show the evolution of $f_{Q}$ as a function of $M_{\rm stream}/M_{\rm h}$. $M_{\rm stream}(z)$ is the halo mass value that marks the transition from hot accretion to cold-in-hot accretion at a given redshift. $M_{\rm stream}/M_{\rm h}$ is a measure of how far into the cold ($M_{\rm stream}>M_{\rm h}$) or hot ($M_{\rm stream}<M_{\rm h}$) accretion regime a group is. This is shown Figure~\ref{fig:qfrac_mstream}, right panel. This plot exhibits even better the trend described above, with quiescent fractions decreasing as $M_{\rm stream}/M_{\rm h}$ values move towards the cold accretion regime. 
{Gray points correspond to the individual groups. The red points are shown only as a visual representation of the underlying trend. We divide the sample into equal-occupancy bins, such that each bin contains the same number of groups; this choice avoids introducing bins with very different statistics in our small sample. For each bin, the member galaxies of all groups in that bin are pooled and treated as a single composite group. We then recompute $f_{\rm Q}$ and its uncertainty using exactly the same procedure described in Section~\ref{section:q_frac}, including the galaxy quiescence probabilities and contamination weights. The binned measurements are not used in either the affine fit or the Spearman rank test, both of which are computed from the individual group measurements.}
To quantify this correlation, we fit the points corresponding to individual groups with an affine function of the form :

\begin{equation}\label{eq:affine}
f(x) = a\cdot x+b \mbox{, where } x={\rm log}_{10}(M_{\rm stream}/M_{\rm h})
\end{equation}

We find a value of $a = -0.13 \pm 0.06$, giving a significance of $2.1\,\sigma$. We also compute the Spearman coefficient corresponding to the individual points and find a value of $r_s = -0.52$, translating to a $p$-value of $p=0.038$ and to a significance of $2.1\,\sigma$ which is coherent with the value derived through the fit of an affine function and with the significance found by \cite{sutanto26}. Both values show this trend is {moderately significant}, with a threshold of 5\% chance probability commonly adopted in the context of astrophysics \citep{feigelson12}. This trend is reasonably expected given the physics at play : the quiescent fractions go from 50\% where gas input happens in hot mode, to 0\% where it happens in cold mode.





\begin{figure}
   \centering
   \includegraphics[scale=0.63]{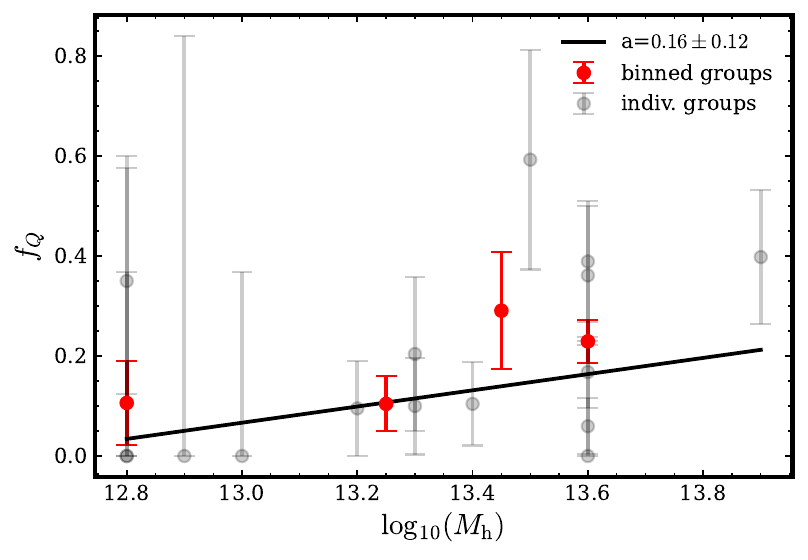}
   \caption{Same figure as the \textit{right} panel of Figure~\ref{fig:qfrac_mstream}, but quiescent fractions are expressed as a function of ${\rm log}_{10} (M_{\rm h})$.}
    \label{fig:qfrac_Mh}
    \end{figure}

{For comparison, we also examine the quiescent fractions directly as a function of halo mass in Figure~\ref{fig:qfrac_Mh}. The quiescent fraction increases with $M_h$. Applying the same analysis as above, we obtain a slope of $a=0.16\pm0.12$, corresponding to $1.4\sigma$, while the Spearman rank test gives a significance of $2.1\sigma$.}

\begin{figure}
   \centering
   \includegraphics[scale=0.63]{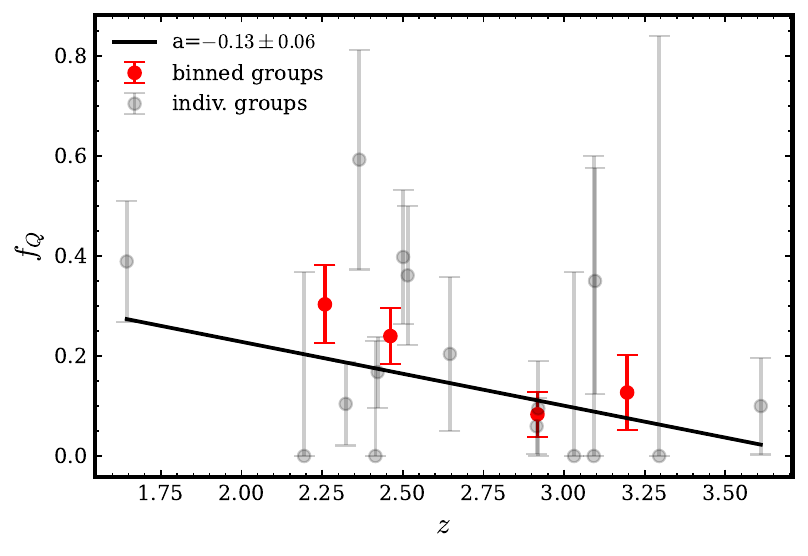}
   \caption{Same figure as the \textit{right} panel of Figure~\ref{fig:qfrac_mstream}, but quiescent fractions are expressed as a function of redshift.}
    \label{fig:qfrac_z}
    \end{figure}

{The statistical strength of the halo-mass correlation is therefore comparable to that obtained with $M_{\rm stream}/M_h$ when considering the rank statistic, although the linear slope is less well constrained.  We similarly show $f_{\rm Q}$ as a function of redshift in Figure~\ref{fig:qfrac_z}. Quiescent fractions increase toward lower redshift, consistent with the evolution measured using the COSMOS-Web galaxy-group catalog \citep{toni25,toni26}. The affine fit gives a significance of $2.0\sigma$, while the Spearman test gives $1.2\sigma$.}

{These three correlations should not be regarded as statistically independent. By construction, a dependence on $M_{\rm stream}/M_h$ necessarily projects in part onto both halo mass and redshift. With the present sample we therefore do not claim that $M_{\rm stream}/M_h$ is statistically demonstrated to be a stronger predictor of $f_{\rm Q}$ than halo mass alone. We nevertheless regard it as the most physically informative parametrization: $M_h$ and $z$ separately describe where a system lies in the halo mass--redshift plane, whereas $M_{\rm stream}/M_h$ measures its position relative to the theoretically predicted, redshift-dependent transition between cold-stream and hot accretion. The correlations with $M_h$ and $z$ can consequently be understood, at least in part, as projections of this two-dimensional accretion framework. The observed correlation alone does not establish causality, and its physical interpretation is investigated further through the galaxy-scale diagnostics presented below.}

\begin{figure*}
   \centering
   \includegraphics[scale=0.7]{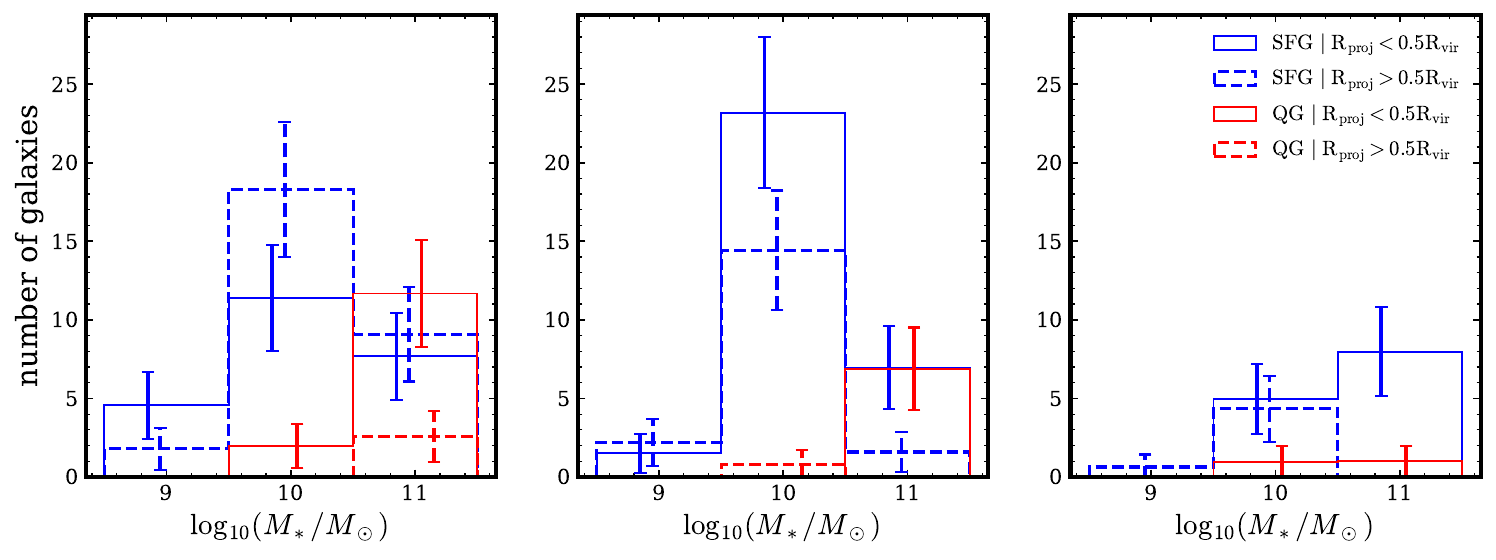}
   \caption{Stellar mass distributions of star-forming (blue) and quiescent (red) galaxies in the inner (full line) and outer (dashed line) regions of the groups. Each panel corresponds to a different bin of  $M_{\rm stream}/M_{\rm h}$ values, grouping structures by their state of gas accretion. Left panel corresponds to groups in the hot accretion regime: $M_{\rm stream}/M_{\rm h}<0.2$. Middle panel corresponds to groups in the intermediate regime: $0.2<M_{\rm stream}/M_{\rm h}<1$. Right panel corresponds to groups in the cold-in-hot accretion regime: $1<M_{\rm stream}/M_{\rm h}$.}
    \label{fig:mass_dist}
    \end{figure*}

\subsection{Mass assembly}\label{results:mass_assembly}

Studying the distribution in stellar mass and group-centric distance of star-forming and quiescent galaxies is an essential point in understanding the mass assembly of clusters and quenching channels of galaxies living in them. Higher mass galaxies form their stellar mass earlier and over shorter time-scales than lower mass galaxies \citep{webb20}, and gain mass through mergers with other galaxies as dictated by hierarchical assembly of structure. Some massive galaxies build up secularly without undergoing any major merger, as seen with the presence of "grand design" spirals, massive regular galaxies at very high redshifts. At different distances to the galaxy groups, processes such as gas replenishment or quenching mechanisms are expected to have varying importance: cosmic web gas accretion happens at the outskirts of groups and the distance at which filaments survive shocks and efficiently penetrate halos as well as the way it trickles down to feed galaxies are open questions and subject to intense study \citep{cornuault18, mandelker20, ramsoy21, dubois21} ; merger activity is more intense moving towards the center of groups \citep{coogan18} where number densities increase and external quenching mechanisms such as galaxy harassment and RAM pressure stripping depend on galaxy masses and on the medium pressure which vary between the inner and outer regions of clusters \citep{xu25, alberts22}. Taken altogether, there are a plethora of reasons to investigate mass/distance distributions of SF/Q galaxies to constrain formation channels.

{Figure~\ref{fig:mass_dist} shows the stellar-mass distributions of star-forming and quiescent galaxies in the inner ($R_{\rm proj}<0.5R_{\rm vir}$) and outer ($R_{\rm proj}>0.5R_{\rm vir}$) regions of the groups. We bin the groups according to $M_{\rm stream}/M_h$, using this quantity as a physically motivated ordering of their expected accretion state, and combine galaxies within each bin to increase the available statistics.} Stellar mass bins in each sub-plot are relatively wide (1 dex) to improve statistics. Results shown in Section~\ref{results:stellar_mass} show that we find a relatively weak dependence of quiescent fractions with stellar mass and as such, using large stellar mass bins is reasonable. Bins are also chosen so that the characteristic mass of $10^{10.5} M_\odot$ sits at the boundary of two bins and not in the middle of one. This characteristic mass has been shown to be important in separating the bimodality of galaxies between the blue/star-forming and red/quiescent galaxy populations at low redshift \citep{baldry04, baldry06} and has been shown to still be relevant when studying this transition at high redshift by \cite{tarrasse25}. In this plot, galaxies are weighted according to the level of contamination from field galaxies in the same way as when computing quiescent fractions in Section~\ref{section:q_frac}.


{The most interesting and robust trend is visible in the highest stellar mass bin. In groups in early evolutionary phase ($1<M_{\rm stream}/M_{\rm h}$, right panel), the high-mass end of the distribution is dominated by star-forming galaxies in the inner parts of the groups, and the second most important population is star-forming galaxies in the outer regions. There is a $4.4\,\sigma$ excess of star-forming galaxies in the inner regions compared to quiescent galaxies. Finally, in the late evolutionary stage ($M_{\rm stream}/M_{\rm h}<0.2$, left panel), quiescent galaxies in the inner region become the prevalent dominant population, outnumbering star-forming galaxies (although with a low significance of $1.6\,\sigma$). Most notably, quiescent galaxies are much more numerous in the inner regions compared to the outskirts, with a $4.4\,\sigma$ excess in the highest mass bin. This is particularly noteworthy when trying to assess the origin of quenching and the feeding of pristine gas to the groups.}

It is particularly interesting to put the results showcased here in contrast with those presented in Section~\ref{results:stellar_mass} which show that stellar mass does not have a prevalent role to explain quiescent fractions. Distance to the groups seems to be a more important factor when investigating the prevalence of quiescent galaxies.

\subsection{Absence of BGG}\label{results:bcg}

\begin{figure}
   \centering
   \includegraphics[scale=0.63]{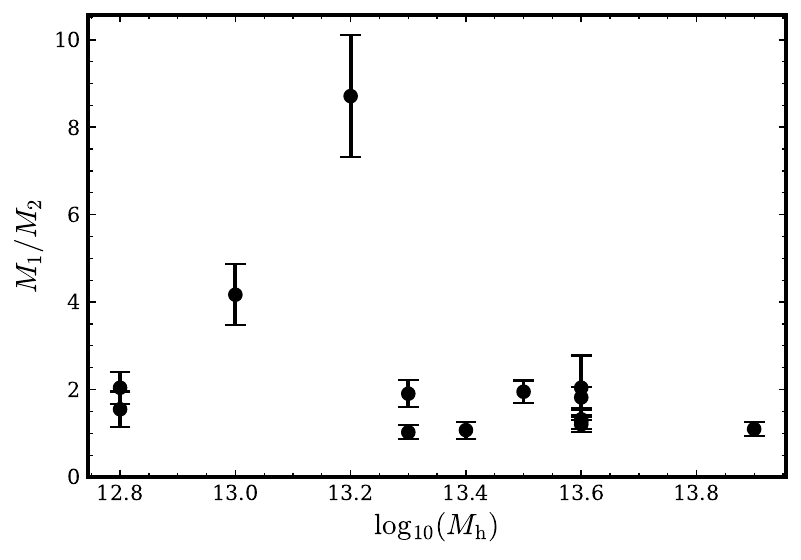}
   \caption{Ratios of the stellar masses of the most to second most massive galaxies in each group, as a function of host halo mass. Most groups have low ratios, showcasing the absence of a clearly established BGG.}
    \label{m12}
    \end{figure}

Witnessing the presence of massive quiescent galaxies in the central regions of the structures we study begs the question of the presence of a clearly established brightest group galaxy (BGG) in our groups. This can be assessed by evaluating the current state of mass assembly of structures. Magnitude gaps are conventionally used as a means to evaluate if systems are dynamically evolved or still forming \citep{jones03, dariush2010, smith2010}. 

The stellar-mass gap between the two most massive galaxies can be used as a high-redshift analogue of the optical magnitude gap commonly adopted at low redshift. We stress, however, that this correspondence should not be understood as a strict transfer of the low-$z$ calibration. At low redshift, magnitude-gap criteria are usually defined in observed optical bands, where luminosity is a reasonable, though imperfect, proxy for stellar mass. By contrast, at $z\sim2$, observed-frame optical magnitudes probe much bluer rest-frame wavelengths and are therefore more sensitive to recent star formation and dust attenuation. For this reason, we adopt the stellar-mass gap directly, which is more closely related to the underlying quantity of interest.

Under the simple assumption that the two galaxies being compared have similar stellar mass-to-light ratios, a low-redshift magnitude gap $\Delta m$ translates into a stellar-mass ratio according to
\begin{equation}
    \frac{M_{*,1}}{M_{*,2}} \approx 10^{0.4\Delta m}
\end{equation}
In this approximation, a commonly used low-redshift threshold of $\Delta m = 2$ corresponds to a stellar-mass ratio of $\sim 6.3$. This conversion should be regarded as heuristic, since it neglects possible variations in stellar population age, dust attenuation, and star-formation history.

Figure \ref{m12} shows the values of the mass ratios of the most massive to second most massive galaxy in each group. In our sample, most groups exhibit much smaller stellar-mass gaps, typically $M_{*,1}/M_{*,2} \lesssim 3$ . Such modest gaps indicate that the most massive galaxy does not yet strongly dominate the massive end of the group population. We therefore interpret these systems as being in an active assembly phase, rather than as dynamically evolved systems analogous to the large-gap groups often identified at low redshift.


\subsection{Stellar mass} \label{results:stellar_mass}

\begin{figure}
   \centering
   \includegraphics[scale=0.63]{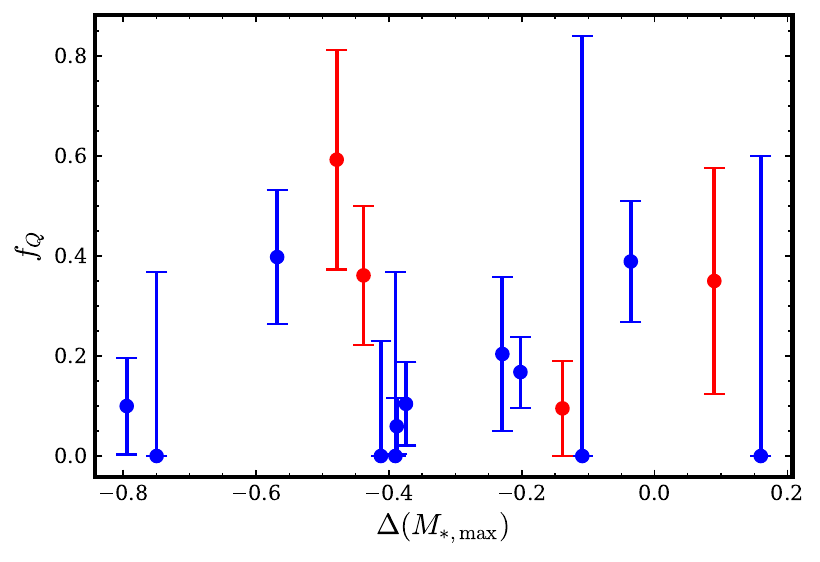}
   \caption{Quiescent fractions of groups as a function of $\Delta M_{*,\max}$ (see text for the definition of this metric). Points are color-coded according to the star-forming (blue) or quiescent (red) status of the most massive member of the group. spearman coef, pvalue and sigma: -0.099 0.72 0.36}
    \label{fig:stellar_mass}
    \end{figure}

To assess whether the group quiescent fraction is more closely linked to halo-scale accretion conditions or to the properties of the dominant galaxy, we define a simple metric based on the stellar-to-halo mass relation (SHMR). For each group, we estimate the average maximum stellar mass expected at its halo mass and redshift using the SHMR from \cite{shmr}, and compute the offset
\begin{equation}
    \Delta M_{*,\max} = {\rm log}_{10} M_{*,\max}^{\mathrm{obs}} - {\rm log}_{10} M_{*,\max}^{\mathrm{SHMR}}(M_{\mathrm{h}},z)
\end{equation}

This quantity measures whether the most massive galaxy in a group is over- or under-massive with regard to what is expected for halos of the same mass and redshift. The values of quiescent fractions as a function of this metric are shown on Figure~\ref{fig:stellar_mass}. 

The motivation for this test is that, if group-wide quenching were primarily driven by internal processes associated with the dominant galaxy \citep[as can be the case in simulations, see for example][]{dashyan19, robson23, sorini22}, one might expect groups hosting an unusually massive central member at fixed halo mass and redshift to display enhanced quiescent fractions. We find no evidence for such a trend: the Spearman rank coefficient between $f_{\mathrm{q}}$ and $\Delta M_{*,\max}$ {translates} to a significance of only $0.36\,\sigma$. This indicates that the stellar-mass excess of the dominant galaxy is not a useful predictor of the group quiescent fraction in our sample.

{We note that $\Delta M_{\ast,\max}$ is predominantly negative in our sample. This may partly reflect the fact that the adopted halo masses are group-scale estimates, while the SHMR comparison refers to the stellar mass expected for a single dominant central galaxy. It may also be physically meaningful: most of our groups lack a clearly dominant BGG and contain several comparably massive galaxies, consistent with ongoing assembly. Their stellar mass may therefore still be distributed among several massive members rather than concentrated in one central object. We consequently use $\Delta M_{\ast,\max}$ primarily as a relative indicator across groups, rather than interpreting its absolute zero point.}

\begin{figure}
   \centering
   \includegraphics[scale=0.6]{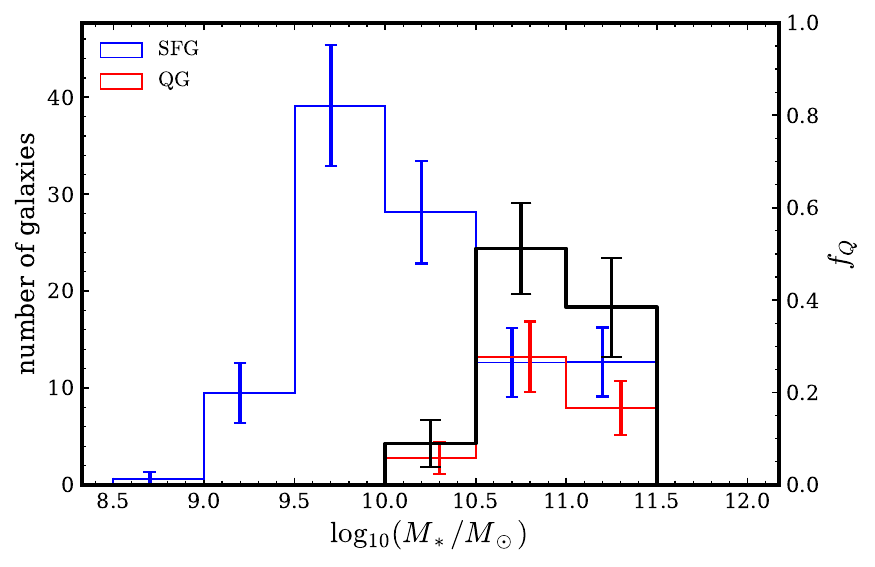}
   \caption{Mass functions of quiescent (red) and star-forming galaxies (blue) in the inner regions of groups undergoing hot accretion and corresponding quiescent fractions (black). There is a decrease of quiescent fractions at the highest stellar masses. Galaxies are weighted according to their corresponding level of sample contamination, as detailed in Section~\ref{results:mass_assembly}.}
    \label{fig:stellar_mass_functions}
    \end{figure}
    
To further probe the role of stellar mass, we plot stellar mass distributions of galaxies in the hot accretion mode (i.e. with $M_{\rm stream}/M_{\rm h}<1$). We group together all groups undergoing hot accretion and plot stellar mass distributions of quiescent and star-forming galaxies, as well as the corresponding quiescent fractions. This is shown on Figure~\ref{fig:stellar_mass_functions}. Galaxies are weighted according to the level of contamination from field galaxies as in Figure~\ref{fig:mass_dist}.
It is visible that the fraction of quiescent galaxies first increases suddenly from the $10^{10}M_\odot<M_*<10^{10.5}M_\odot$ bin to the $10^{10.5}M_\odot<M_*<10^{11}M_\odot$, corresponding to the transitional appearance of quiescent galaxies. But from the $10^{10.5}M_\odot<M_*<10^{11}M_\odot$ bin to the $10^{11}M_\odot<M_*<10^{11.5}M_\odot$ bin, there is no increase in the quiescent fraction. Within the uncertainties, the fraction even seems to decrease. There seems to be no monotonic increase of the quiescent fraction with stellar mass. This trend remains, although with lower statistics, when using finer stellar mass bins to have 3 bins instead of 2 above $10^{10.5}M_\odot$. This trend is at odds with low redshift expectations where quiescent fractions increase exponentially with stellar mass \citep{peng10, davies16} but consistent with results obtained at similar redshift, where this fraction starts rising around $10^{10.5}M_\odot$ and flattening around $10^{11.5}M_\odot$ \citep{sherman20}. This further supports the idea that internal processes do not play a major role in quenching the galaxies in the groups we observed as we would expect a rise in quiescent fractions with stellar mass if it was the case. Altogether, our results point to a greater role of environment over internal processes to explain the population of quiescent galaxies in the structures we study.

\subsection{Sensitivity to choices of cuts and study limitations}\label{section:limitations}

The results showcased in this study are robust to contamination selection and alterations in the criteria used to classify galaxies as quiescent. All results based on quiescent fractions or stellar mass distribution account for field contamination through the weighting of galaxies. Altering the quiescent criteria in a more stringent way also preserves these trends. Our way of computing quiescent fractions is also robust to the definition of quiescence, as galaxies are associated a $p$-value depending on the overlap of the posterior of their physical properties with the criteria used, meaning that varying such criteria would not mean a galaxy close to the threshold would suddenly switch classification when varying its definition.


{The adopted maximum projected radius for the member search is another methodological choice that can affect the selected galaxy population. Because these high-redshift groups may not yet be fully virialized, there are two reasonable approaches. One is to include galaxies beyond $R_{\rm vir}$ that may subsequently be accreted by the halo; this would also provide greater sensitivity to pre-processing in the group outskirts. The more conservative alternative is to restrict the analysis to galaxies currently within $R_{\rm vir}$, since the physical conditions outside the virial radius may differ substantially from those in the group interior. We adopt the latter approach.}

We chose to go with the second approach. We had initially chosen to search for members slightly beyond their virial radius ($1.3 R_{\rm vir})$, which is common in other works studying the same type of objects we study here, see for example \cite{esposito25}. However, at least one of our groups has a known satellite at such close distance that the satellite member galaxies would fall into the main halo's reach of membership using this initial cut. To avoid confusion and double counting, we chose to shrink our area of search to strictly within the virial radius of the groups. We emphasize that without prior knowledge of the presence of this satellite, choosing to go beyond the virial radius of structures would have been a reasonable choice as well but would have changed in some aspects the interpretation of the results. Using a more limited area of search most likely limits the number of quiescent galaxies undergoing pre-processing in the outskirts, like the many found by \cite{sillassen26} in the Cosmic Vine \citep{jin24}.

\section{Discussion} \label{section:discussion}

\subsection{Internal quenching}

Quenching is traditionally viewed as a phenomenon happening through two channels: internal and environmental processes \citep{man18, alberts22}. 
Quenching by internal processes is usually assessed using stellar mass as a proxy. This makes sense given that the rate of supernovae should be related to the stellar mass of a galaxy and given the stellar to black hole mass relation \citep{reines15} which implies that their feedback should scale with stellar mass. However, linking AGN feedback to galaxy quenching is not straightforward.

Some studies find AGN to be unambiguously associated with quiescent galaxies \citep{olsen13} while others on the contrary find AGN presence to be mostly associated with star-forming galaxies \citep{coleman22}, even in simulations \citep{ward22}, or simply not over-represented in quiescent galaxies \citep{almaini25}. \cite{silverman25} find that even the most luminous QSO in the universe at z=2 does not seem to suffice to quench star formation in their hosts. AGN activity, being a phenomenon regulated by gas input just like star-formation, might be co-occurring with quenching rather than causing it \citep{mullaney12, kubo22}.

Although it is not the main focus of this study and the data we use allows limited investigation in assessing AGN activity, we have tried investigating the role of internal processes in quenching galaxies in our groups. This was done by looking at the evolution of the quiescent fraction with stellar mass and the correlation between quiescent fractions in groups with how over-massive the most massive galaxy of the group is compared to what is expected at the same halo mass and redshift. In both cases, our results seemed to disfavor internal processes as a compelling explanation. While these results do not rule out that internal processes may play a part in quenching of individual galaxies, they suggest that they are unlikely to be the main reason to explain the presence of a group-wide quiescent population in our sample.


\subsection{Environmental processes}

In recent years, several studies have found evidence for a lack of environmental quenching at high redshifts \citep{huang25, shi25, pan25}. However, most of these studies base their definition of high-density environments at the very large scale, studying either the large scale structure through overdensity mapping or single structures that are protoclusters, several Mpc in size, where physics cannot be expected to be the same as those of more compact groups. Protoclusters are inherently defined as regions that will collapse as an overdensity in the future, and while studying the physics of these regions is meaningful, it cannot serve as a basis to discard environmental effects as a whole at high redshift. 

Many studies find environmental processes to still play a major role in shaping galaxies at high redshift, mainly through merger processes affecting the gas content and star-formation properties of galaxies \citep{tan24, puglisi21, coogan18, xu25}. These processes can contribute to quench galaxies present in the structures we observe as well by accelerating their evolution or depleting them of their gas.

\subsection{Cosmic starvation/hot accretion quenching}

On top of the increased rate of interactions between galaxies, over-dense environments impact them through cosmic web gas accretion and structure virialization. Gas depletion timescales of a few hundred million years for star forming galaxies mean that galaxies need constant refueling to remain star-forming \citep{daddi10, tacconi10, tacchella16}. At the largest scales, this refueling happens through gas accretion coming from cosmic web filaments connecting to halos. However, few studies have investigated the direct connection between the two: a direct link was first suggested by \cite{overzier16} and \cite{behroozi19} probed directly the link between cosmic gas accretion and the properties of galaxies being fed the gas from the perspective of simulations.

{Although these flows are extremely difficult to image directly, many studies point to their role in shaping galaxy evolution. The cold-to-hot transition offers a physical explanation for this role: gravitational shock heating can suppress the fresh cold-gas supply that sustains star formation. We find a moderate ($2.1\sigma$) correlation between $f_{\rm Q}$ and $M_{\rm stream}/M_h$. As discussed in Section~\ref{results:cold_accretion}, the present sample does not establish that this parameter is a stronger statistical predictor than $M_h$ alone. Comparing these correlated parameters and testing alternative explanations nevertheless helps us assess the physical origin of the trend. Together with the radial and stellar-population diagnostics, the results lead us to interpret the accretion-state transition as a contributor to quenching in these groups, although they do not uniquely isolate its effect from those of halo mass and redshift. $M_{\rm stream}/M_h$ has the advantage of encoding the expected redshift-dependent accretion transition and the growth and maturation of groups over time, rather than treating halo mass or redshift separately. We therefore use it as the physically motivated framework for interpreting our results.}


{The same global dependence on the expected accretion regime is recovered by \citet{sutanto26} in their much larger sample of photometrically selected overdensities. Their result provides an independent large-statistics confirmation that the correlation is not restricted to the comparatively small number of spectroscopically confirmed systems studied here. The analysis presented in this work goes beyond this global correlation by investigating the galaxy populations responsible for it. In particular, we use the stellar-mass and radial distributions of quiescent galaxies to constrain how quenching proceeds within the groups, test whether stellar mass or the properties of the dominant galaxy provide alternative explanations for the group-wide quiescent fractions, and examine the sensitivity of these results to field contamination, quiescence classification, radial selection, and SED-fitting assumptions.}

In detail, we find all quiescent galaxies to be massive ($M_*>10^{10.5}M_\odot$) and in the central parts of groups with a $4.4\sigma$ excess of massive quiescent galaxies in the central parts of evolved groups compared to the outskirts. This is coherent with the picture of progressively virializing groups and disrupting filamentary accretion from the inside-out. We find distance to the group to be the most important factor when assessing the quiescence of galaxies as opposed to stellar mass. The fact that these galaxies are overwhelmingly massive most likely reflects the fact that they started forming earlier than less massive galaxies \citep{webb20} and have exhausted their cool gas reservoir first.


Recent studies making use of the incredible sensitivity in the infrared of the JWST find a relative lack of low mass quiescent galaxies at the redshifts we probe, showing that the picture put forward in this study likely does not suffer from the shallower depth of the COSMOS2020 dataset \citep{shuntov25}. 
{Taken together, these observations are consistent with quenching in high-redshift groups being influenced by cosmic starvation, accelerated evolution, and increased merger rates associated with the overdense environment \citep{wetzel13, zabludoff98}.}

The picture is however likely more complicated than what might seem.
\cite{guo25} recently discovered an unambiguously quiescent BGG located in the very center of a giant Ly$\alpha$ nebula, RO-1001-Sat.  \cite{wangw26} presents a similar case, the proto-cluster MQN01, see also \cite{kalita22}. The presence of quiescent galaxies in the middle of a giant gas reservoir shows that having cold gas is not sufficient for enabling star-formation, as this gas can be rendered inefficient either by AGN feedback, morphological quenching or other processes. The BGG in \cite{guo25} sits exactly at the position of peak luminosity of the nebula which corresponds to the bottom of the gravitational well and there are conditions for the cold gas to flow inside it, but no star-formation is observed. The hosting group classifies as a fossil group and as such might be dynamically more evolved, it might be a prime illustration of the onset of hot accretion but it still contains lots of cold gas. The center of RO-1001-Sat may have just entered the state in which cold filaments get disrupted and filamentary accretion becomes inefficient. Its place in the halo mass -- redshift diagram seem to support this interpretation, being right at the boundary between the hot and cold-in-hot accretion regime.

\subsection{Survival of filaments}

The question of the survival of filaments infalling on groups and galaxies is crucial in such a context. Several works have investigated this point and come to different conclusions: \cite{danovich15} for example have found that filaments penetrate halos and feed galaxies efficiently, while on the other hand, \cite{medlock25} find that filaments around groups do not survive interactions with the CGM heated and perturbed by feedback and are inefficient at feeding galaxies. These studies are sensitive to the choices of feedback implementations and numerical resolution and the question of survival of filaments remains open. 

{The high quiescent fractions observed in groups located in the hot-accretion region of the halo mass--redshift diagram (Figure~\ref{fig:qfrac_mstream}) are consistent with a scenario in which the bulk of cold filamentary inflow no longer survives efficiently within the virial radius of mature groups.} The residual accretion may still happen as visible by the presence of star-forming galaxies in the inner regions of these mature groups undergoing hot accretion (Figure~\ref{fig:mass_dist}). Galaxies may fall in groups with their gas reservoir allowing them to sustain star-formation for several billion years even though the group is undergoing starvation overall \cite{fossati17}. The occasional presence of quiescent galaxies in groups close to the boundary between cold and hot accretion or even within the cold accretion regime is not incompatible with the cold accretion paradigm. These quiescent galaxies may have underdone cosmic web detachment, and even though their host halo still accretes cold gas, individual galaxies may fall out of line with the filaments connecting to its parent halo and starve as a result \citep{aragoncalvo19}.

\subsection{Assembly of clusters at cosmic noon}

Analysis of the mass distributions of galaxies in each group seem to indicate that in general, there is not a clear established BGG. Two of our groups meet the criteria for fossil groups, but their BGG is star-forming, an indication that even though they are more ancient, they are still actively forming \citep{einasto22}.

The rest of our sample have several most massive members of comparable masses, often times at similar projected distances to the group. This indicates that most of the structures we are observing are undergoing assembly and actively merging with other groups of similar mass \citep{einasto12}. 

Figure \ref{m12} along with the results shown in section \ref{results:bcg} show that we are witnessing groups well in the process of assembling mass. In most cases, groups don't have a clear BGG in place with several star-forming or quiescent galaxies of comparable masses close to the center of the structures and low mass gaps between the most massive galaxies. Hierarchical structure formation would suggest that each of the most massive galaxies in the structure can be seen as a tracer of the parent structure it belonged to before merging, which would mean that most of the structures we observe result from relatively recent mergers from parent groups of comparable masses.

\section{Summary and conclusions}

In this study, we have presented analysis of the quiescent population of a sample of spectroscopically confirmed high-z galaxy groups and its relation with the properties of their host halos. We have developed a new method to select member galaxies, a procedure to fit their SED with the physically appropriate M13 stellar templates and classify them as quiescent according to several criteria based on their physical properties, while propagating statistical and classification uncertainties consistently. We then investigated the relation between the distribution of quiescent galaxies and the properties of their host halo. Our main findings are:

\begin{itemize}
    \item Stellar mass does not seem to be the main driver of quiescence in groups: quiescent fractions in groups undergoing hot accretion decrease at the highest stellar masses and quiescent fractions in groups do not correlate with how over-massive the most massive member is
    \item The quiescent fractions of groups show the strongest correlation with the cold/hot state of the gas being accreted on groups; they vary from 50\% in groups in the hot accretion regime and decrease to 0\% in groups with higher $M_{\rm stream}/M_{\rm h}$ values
    \item Quiescent galaxies are overwhelmingly massive ($M_*>10^{10.5}M_\odot$) and located in the inner parts of mature groups undergoing hot accretion ; there is a $4.4\sigma$ excess of massive quiescent galaxies in the inner regions compared to the outskirts of these groups 
    \item These results are robust to field contamination, to variations of the quiescent criteria and maximum scanning radius used
    \item {The quiescent fraction also correlates with halo mass, with a rank-correlation significance comparable to that obtained for $M_{\rm stream}/M_h$. This is not an independent correlation, since $M_{\rm stream}/M_h$ explicitly depends on $M_h$ and redshift. We favor the accretion-state parametrization as the physical framework for interpreting the trends because it places each halo relative to the theoretically predicted transition in gas-supply mode. }
\end{itemize}


{Taken together, these results are consistent with cold-to-hot accretion playing a role in the onset of quiescence in the groups we study. The radial distribution of the quiescent population is particularly suggestive of an inside-out progression, as expected if the disruption of filamentary gas supply becomes increasingly effective as the intra-group medium is shock heated. The present data establish correlations rather than a unique causal connection, but the agreement between the global accretion-state trend, the spatial distribution of quiescent galaxies, and the tests of alternative internal explanations makes cosmic starvation a compelling interpretation of the observations.}


Obtaining spectroscopic confirmation of membership would allow to better constrain statistics derived on groups especially for low mass galaxies where contamination is more important. Other studies using deeper data do not seem to uncover an unknown population of quiescent galaxies and as such we do not believe the interpretations of our study to be limited by the lack of depth of COSMOS2020. Although a sample of 16 spectroscopically confirmed groups is relatively large and allows to probe the halo parameter space to a satisfying degree, expanding this sample and deepening our understanding of their member galaxies would allow to refine our understanding of the cold to hot accretion transition and its impact on the onset of quiescence.

\begin{acknowledgements}
This project has received funding from the European Union’s Horizon 2020 research and innovation programme under the Marie Skłodowska-Curie grant agreement No 101148925. R. G. acknowledges funding from project ANID Fondecyt 1231661. S.L. acknowledges the support from the National Natural Science Foundation of China (No. 12503011) and the Key Laboratory of Modern Astronomy and Astrophysics (Nanjing University) by the Ministry of Education.
\end{acknowledgements}

\bibliographystyle{aa} 
\bibliography{biblio.bib} 


\begin{appendix} 

\section{SED fitting systematics}

\begin{table*}[t]
\caption{Grids of values used to determine if the fit is numerically converged by comparing results of the 'default' and 'fine' grid runs.}
\label{tab:grids}
\centering
\begin{tabular}{l ccc ccc ccc}
\toprule
 & \multicolumn{3}{c}{${\rm log}_{10} (t)$} & \multicolumn{3}{c}{${\rm log}_{10} (\tau)$} & \multicolumn{3}{c}{$\rm A_v$} \\
\cmidrule(lr){2-4} \cmidrule(lr){5-7} \cmidrule(lr){8-10}
Grid     & min & max  & step &  min & max  & step & min & max  & step \\
\midrule
Default & 8 & 10  & 0.1 &  8 & 11  & 0.1 & 0 & 5  & 0.02 \\
Fine & 8 & 10  & 0.05 &  8 & 11  & 0.05 & 0 & 5  & 0.01 \\
\bottomrule
\end{tabular}
\end{table*}

A major drawback of modern SED fitting softwares is their lack of flexibility when it comes to stellar libraries. Most of them stick to the most commonly used ones and it is difficult to import custom stellar libraries into them. Using realistic templates is just as important as using cutting edge SED fitting methods, if not more important. Fast++ thankfully offers the possibility of easily importing new stellar templates. The Bruzual and Charlot 2003 templates are used almost exclusively in the field to study galaxy physics. However, the question of the contribution of thermally-pulsating asymptotic giant branch (TP-AGB) stars to the overall energy distribution of galaxies has been an open question for decades and each set of stellar templates have different implementations for that phase resulting in varying energetics and contributions to the overall balance. Recently, \cite{lu24} has focused on studying high-resolution spectra of quiescent galaxies showing important TP-AGB spectral features and has found that the Maraston 2013 templates show the better agreement and capture the most spectral features out of all templates tested. As such, using SED fitting software allowing the use of the M13 was a priority.

Interestingly, when comparing the SFRs derived using the M13 templates to the ones obtained with the BC03 ones, we find that they are 0.6 dex higher on average which perfectly offsets what \cite{pacifici2015} find to be underestimated by classical SED fitting set-ups. \\


We present here the analysis conducted to ensure the robustness of the fitting results with regard to the choice of grid of fitted values and stellar templates. 

\subsection{Grid}\label{appendix}

Table \ref{tab:grids} shows the two grids of parameters tested to ensure the fit reached convergence. The fine grid has a step twice as small as the default one to ensure the minimum $\chi^2$ isn't just a local minimum. 

\begin{figure*}
  \centering
  \includegraphics[scale=0.5]{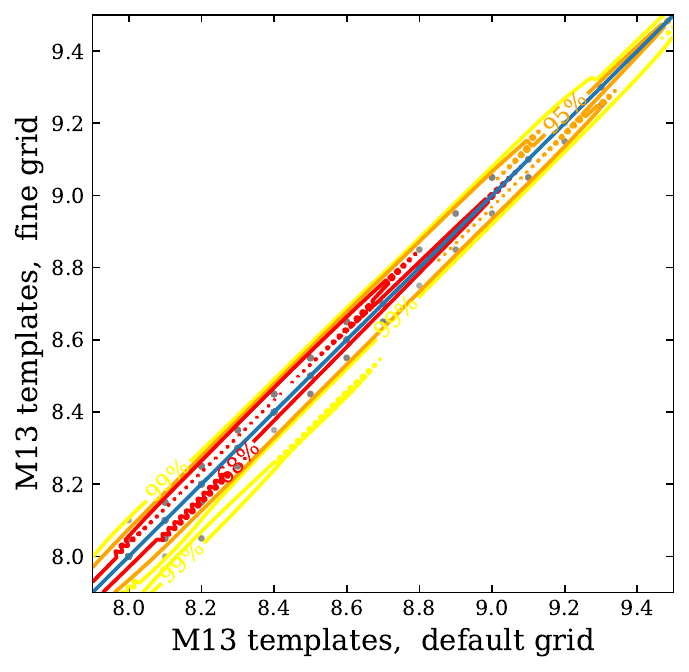}
  \includegraphics[scale=0.5]{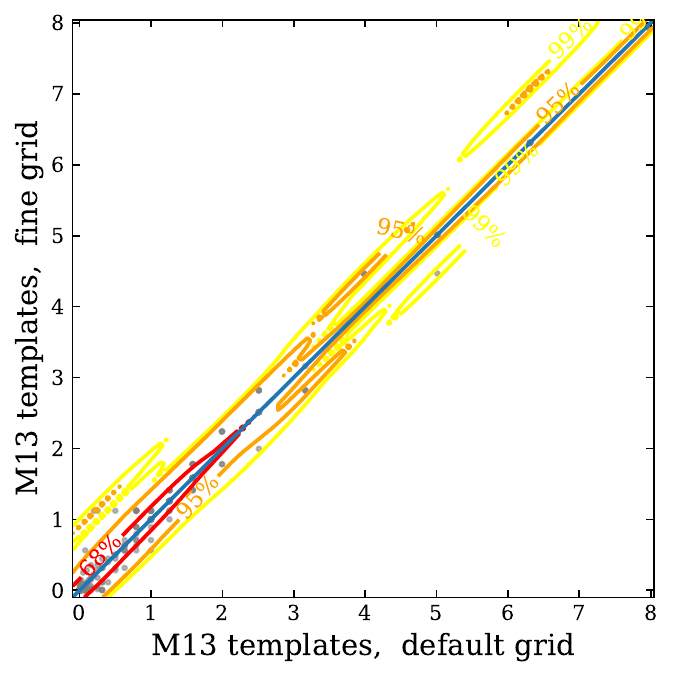}
  \includegraphics[scale=0.5]{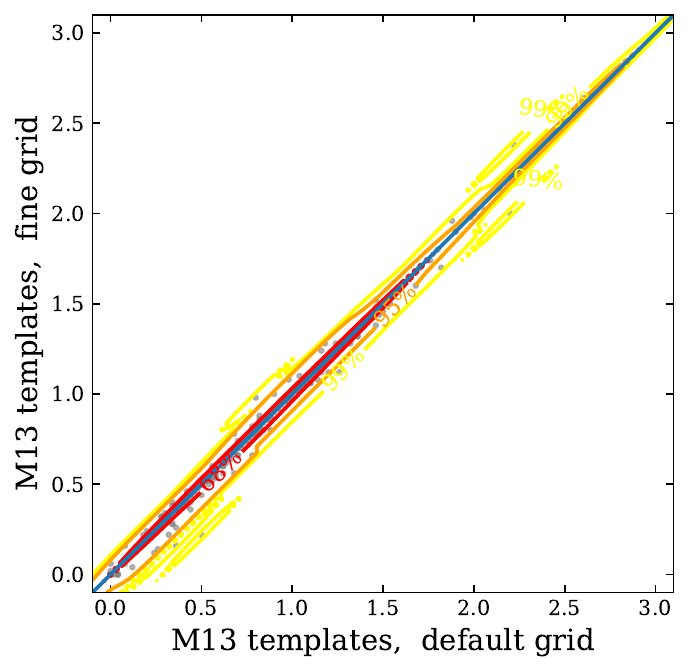}
  \caption{Figure comparing the values of best fit parameters when varying the coarseness of the grid. Red, orange and yellow contours correspond to the levels encapsulating $68\%$, $95\%$ and $99\%$ of the data points respectively. \textit{Left} panel shows values for ${\rm log}_{10}(t)$ with ages in years, \textit{middle} panel shows values for $t/\tau$, \textit{right} panel shows values for $\rm A_v$.}
    \label{fig:grids}
    \end{figure*}

Figure \ref{fig:grids} shows the distributions of best fit parameters for the age of the stellar populations, the $t/\tau$ value used in the SFH and the value of dust attenuation. All three of them show tight 1:1 correlations with low scatter (~0.1 dex) and few outliers. Hence, it is safe to say the fit is numerically converged and the solution found with the default grid corresponds to the true minimum of the problem. 

\subsection{Templates}

\begin{figure*}
  \centering
  \includegraphics[scale=0.5]{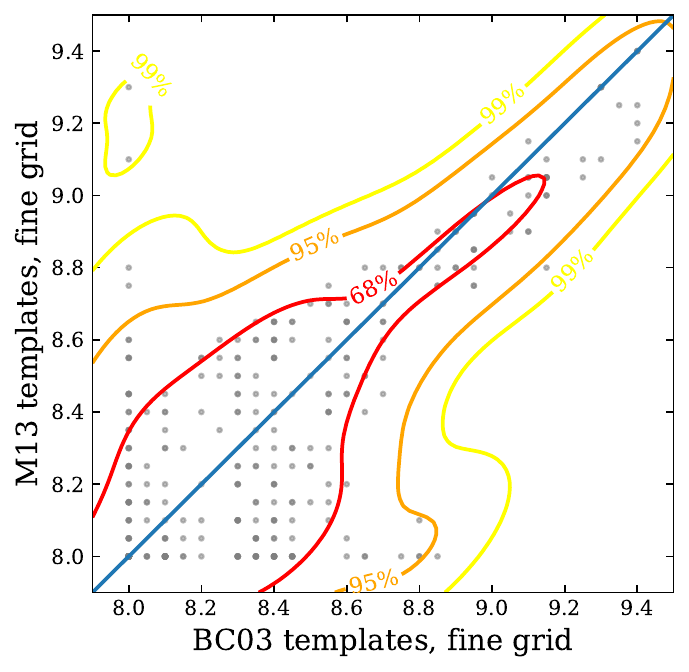}
  \includegraphics[scale=0.5]{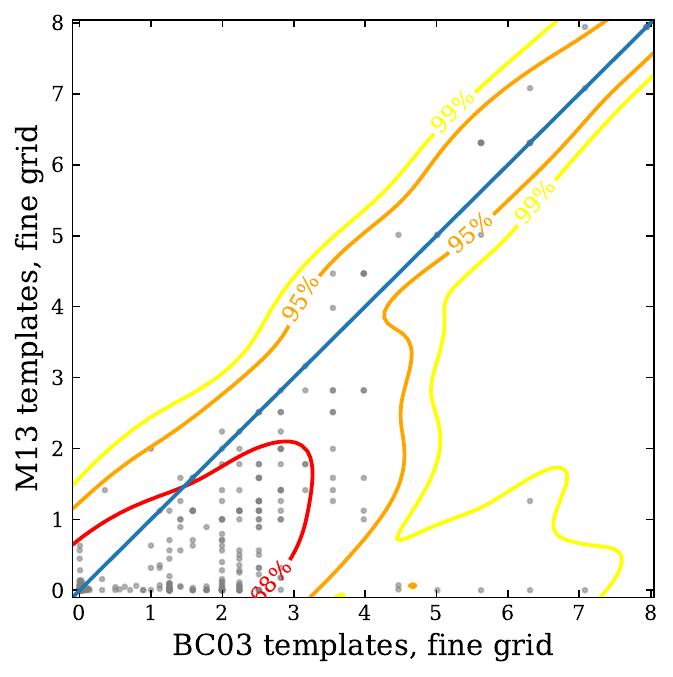}
  \includegraphics[scale=0.5]{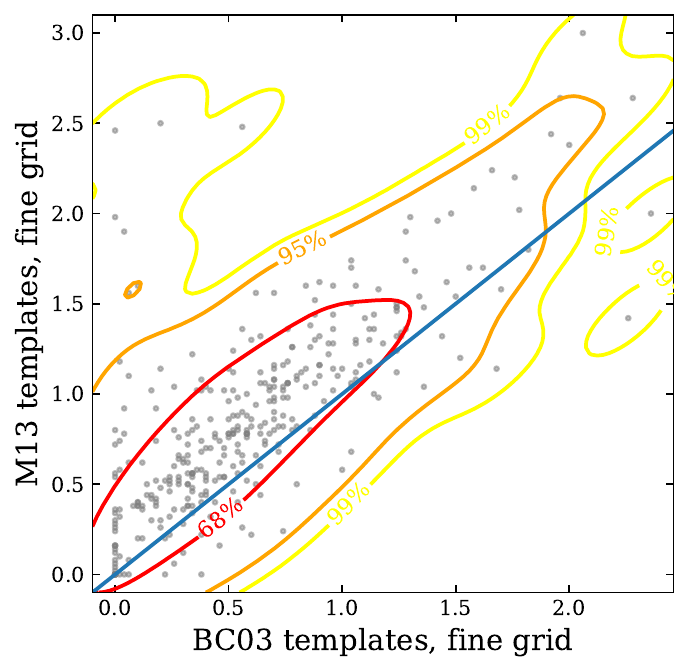}
  \caption{Same as Figure~\ref{fig:grids} but when varying the templates used instead of the grid.}
    \label{fig:templates_param}
    \end{figure*}

To explore the impact of choosing the M13 templates over the more commonly used BC03 ones and to allow for comparison with the rest of the literature, we show in Figure~\ref{fig:templates_param} the systematic differences between the best-fit values when using the two different set of templates. Ages don't show significant bias across templates, scatter is low for lower ages (${\rm log}_{10} (t)<8.8$) typically 0.3 dex and decreases at higher values, albeit with less sampling. $t/\tau$ deviate slightly across templates, with M13 values being typically smaller than the BC03 counterparts at the low-end. Fortunately, most of the galaxies with $t/\tau>3$ have concordant values across template, meaning that the quiescent classification holds across templates. Finally, Av values seem relatively insensitive to the choice of templates, following a close 1:1 relation with low scatter (~0.5 dex).

\begin{figure*}
  \centering
  \includegraphics[scale=0.5]{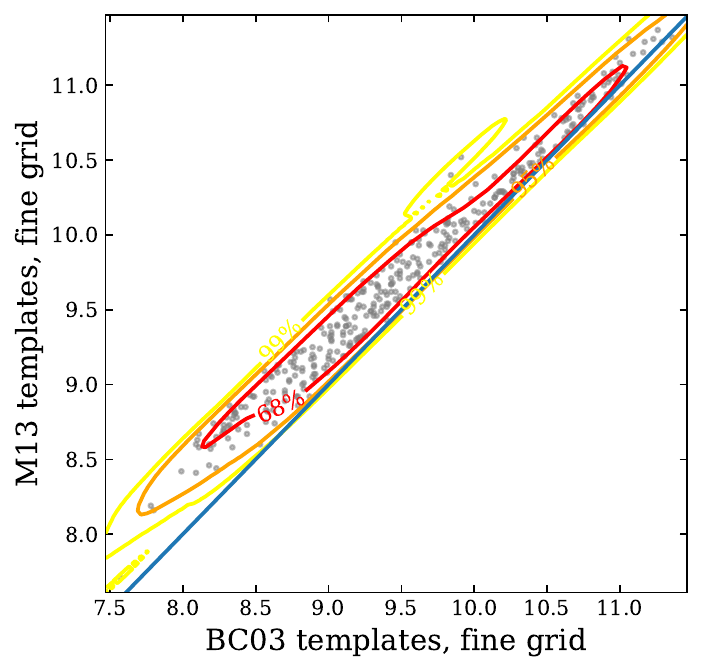}
  \includegraphics[scale=0.5]{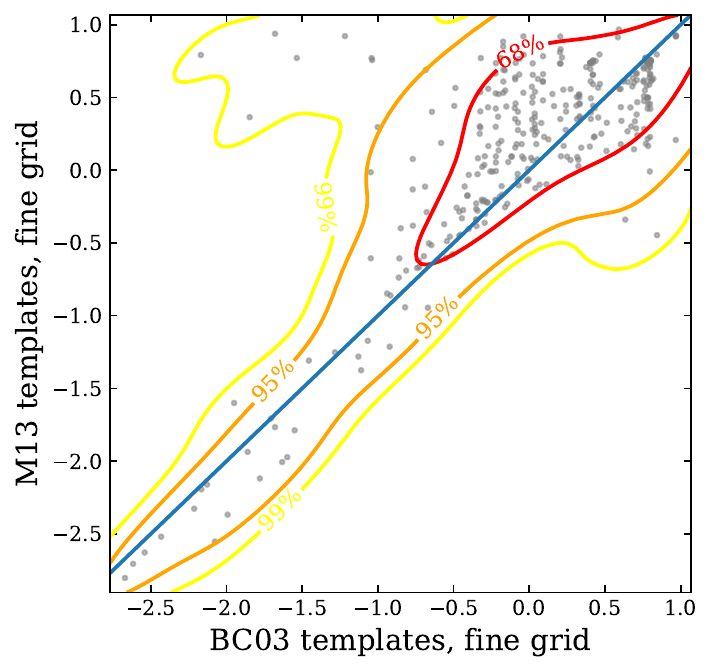}
  \includegraphics[scale=0.5]{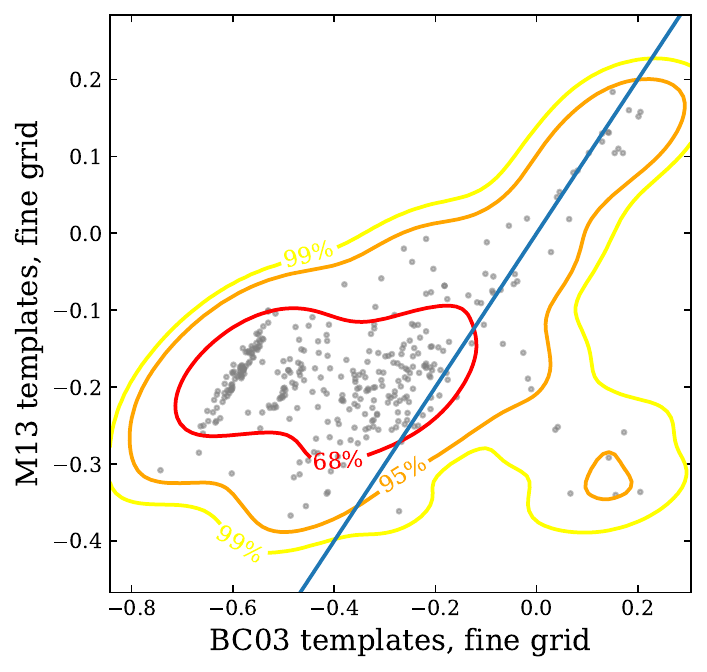}
  \caption{Same as Figure~\ref{fig:templates_param} but showing the values for $\rm{log_{10} (M_*/M_\odot)}$ on the \textit{left} panel, distance to the main sequence on the \textit{middle} panel and distance to the border in the UVJ diagram in the \textit{right} panel.} 
    \label{fig:templates_prop}
    \end{figure*}

Furthermore, Figure~\ref{fig:templates_prop} shows the same for galaxy properties used to classify them as quiescent or star-forming. Masses don't suffer from much bias related to template selection, scatter is low (~0.15 dex) but the relationship shows slightly higher masses predicted by the M13 templates. In the range of interest for our study, namely under O.6 dex under the SFMS, the values of distance to the main sequence are tightly correlated and show no bias one way or another. There is more spread in the SF region, likely tied to the relative unfitness of the declining SFH in this regime. Finally, UVJ distances are in agreement between the to sets of templates in the from about -0.1 mag (meaning 0.1 mag towards the star forming region of the UVJ diagram) and further in the quiescent galaxy region, which is the most important for our work. There is significantly more spread in the SF area, with a large span of values in the BC03 predicted UVJ distances all equivalent toa much tighter area in the M13 ones, which seems to indicate that the M13 templates are more degenerate in this regime.

\subsection{Assessing fitting quality}

\begin{figure}
  \centering
  \includegraphics[scale=0.5]{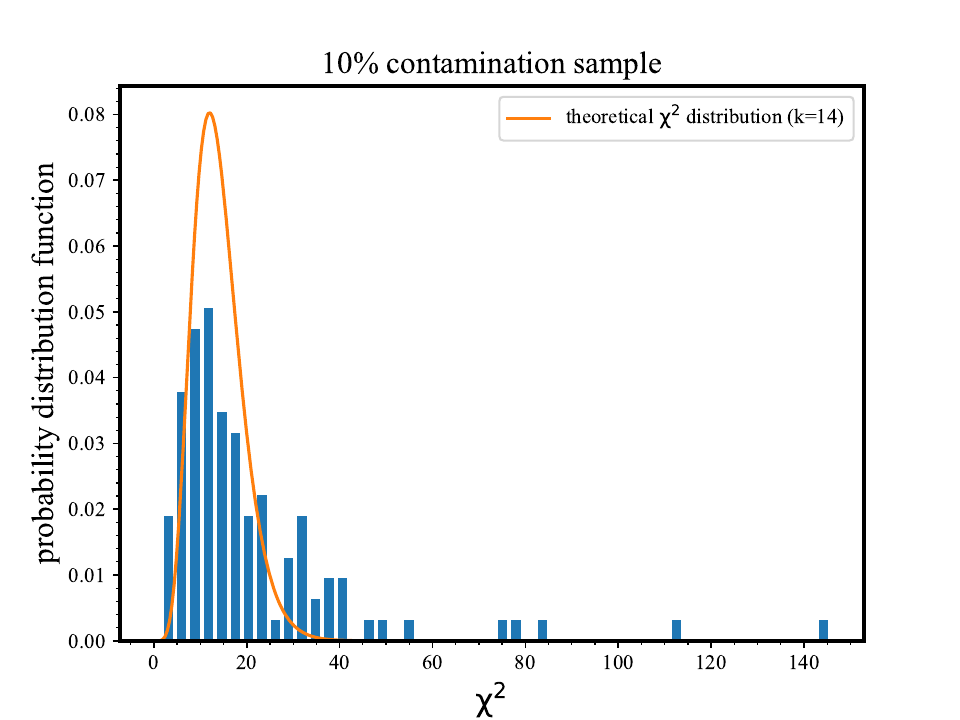}
  \includegraphics[scale=0.5]{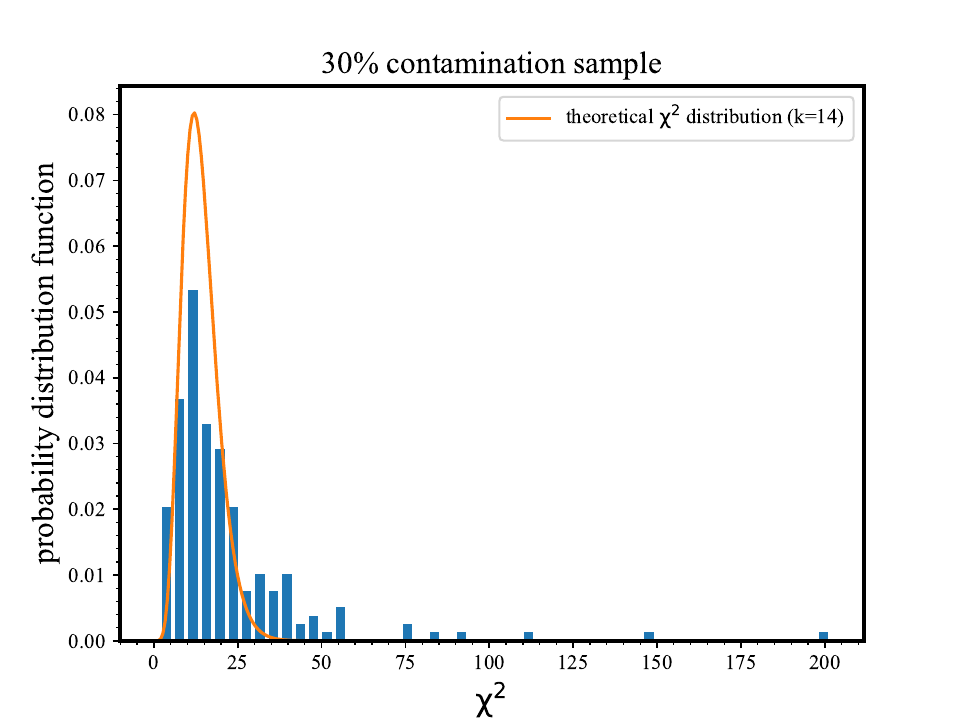}
  \caption{$\chi^2$ distributions in the 10\% (\textit{top} panel) and 30\% (\textit{bottom} panel) as well as theoretical $\chi^2$ distributions.}
    \label{fig:chi2}
    \end{figure}
    
To assess the quality of the fitting routine developed, we check the $\chi^2$ distributions among the samples at 10\% and 30\% contamination. Both show good agreement with the theoretical probability distribution function, with notable excess around $20<\chi^2<40$. There are also a few cases of fits with very high $\chi^2$, most of which are due to close undetected neighbors polluting the photometry or to inhomogeneous photometry and errors across bands. Some bands used for the fitting are very close wavelength-wise but have differing fluxes due to the fact that they come from different facilities (for example the HSC-y and UVISTA-Y filters which are overlapping at some point). Manual visual inspection of the observed and best-fit SED as well as the cutouts from the bands used around the galaxy in question still allows to ascertain that in a large fraction of the worse fits in our sample, the physical information extracted from the fits is still relevant to what is visible from the cutouts.
    




\end{appendix}

\end{document}